\documentclass[%
preprintnumbers,
amsmath,
amssymb,
aps,
]{revtex4-2}

\usepackage{graphicx}
\usepackage{subfigure}
\usepackage{comment}

\usepackage{dcolumn}
\usepackage{bm}

\usepackage{tikz}
\DeclareRobustCommand\full  {\tikz[baseline=-0.6ex]\draw[thick] (0,0)--(0.5,0);}
\DeclareRobustCommand\dashed{\tikz[baseline=-0.6ex]\draw[thick,dashed] (0,0)--(0.54,0);}
\DeclareRobustCommand\chain {\tikz[baseline=-0.6ex]\draw[thick,dash dot dot] (0,0)--(0.5,0);}

\usepackage{color}
\usepackage{tikz}
\usetikzlibrary{shapes}
\newcommand{\filledbox}{\raisebox{0.5pt}{\tikz{\node[draw, scale=0.8, rectangle, fill=black!100!](){};}}}
\newcommand{\filledboxred}{\raisebox{0.5pt}{\tikz{\node[draw, scale=0.8, rectangle, fill=red!100!](){};}}}
\newcommand{\filledboxgreen}{\raisebox{0.5pt}{\tikz{\node[draw, scale=0.8, rectangle, fill=green!100!](){};}}}

\newcommand{\filleddiamondblue}{\raisebox{0pt}{\tikz{\node[draw,scale=0.5,diamond,fill=blue!100!](){};}}}

\newcommand{\filledtriangleblack}{\raisebox{0.6pt}{\tikz{\node[draw,scale=0.4,regular polygon, regular polygon sides=3,fill=black!100!,rotate=0](){};}}}
\newcommand{\filledtriangleblue}{\raisebox{0.6pt}{\tikz{\node[draw,scale=0.4,regular polygon, regular polygon sides=3,fill=blue!100!,rotate=0](){};}}}
\newcommand{\filledtrianglered}{\raisebox{0.6pt}{\tikz{\node[draw,scale=0.4,regular polygon, regular polygon sides=3,fill=red!100!,rotate=0](){};}}}
\newcommand{\filledtrianglegreen}{\raisebox{0.6pt}{\tikz{\node[draw,scale=0.4,regular polygon, regular polygon sides=3,fill=green!100!,rotate=0](){};}}}
\newcommand{\filledtriangleyellow}{\raisebox{0.6pt}{\tikz{\node[draw,scale=0.4,regular polygon, regular polygon sides=3,fill=yellow!100!,rotate=0](){};}}}
\newcommand{\filledgradientred}{\raisebox{-1.0pt}{\tikz{\node[draw,scale=0.45,regular polygon, regular polygon sides=3,fill=red!100!,rotate=180](){};}}}
\usepackage{mathtools}

\newcommand{\xx}{\boldsymbol{x}}

\newcommand{\dd}{\text{d}}

\usepackage{enumitem}

\usepackage{soul}

\soulregister\cite7
\soulregister\ref7
\soulregister\emph7

\usepackage{placeins}

\begin{document}

\preprint{}

\title{Flow-acoustic resonance in deep and inclined cavities}

\author{You Wei Ho$^1$}
\email{Corresponding author: ywh1u18@soton.ac.uk}
\author{Jae Wook Kim$^2$}%
\affiliation{%
$^1$ Institute of Sound \& Vibration Research, University of Southampton, Southampton, SO17 1BJ, United Kingdom \\
$^2$ Aeronautics \& Astronautics, University of Southampton, Southampton, SO17 1BJ, United Kingdom
}%

\date{\today}

\begin{abstract}	
This paper presents numerical investigations of flow-acoustic resonances in deep and inclined cavities using wall-resolved large-eddy simulations. The cavity geometry considered has a fixed aspect ratio of $D/L = 2.632$ and is subjected to two Mach numbers of $0.2$ and $0.3$ at three different angles of inclination ($\alpha=30^{\circ}$, $60^{\circ}$, and $90^{\circ}$). Fully turbulent boundary layers generated from independent precursor simulations are employed upstream of the cavities. The simulation results show significant differences in aeroacoustic response between inclined and orthogonal cavities, particularly at $M_{\infty} = 0.3$, where the inclined cavities exhibit stronger resonances (by more than 20 dB) at a lower peak frequency ($St=0.276$) than the orthogonal cavity, whose peak occurs at $St=0.849$. Acoustic modal analysis identifies these frequencies as the 1st and 2nd eigenmodes, respectively. Further analysis shows that the difference in mode selection is linked to the hydrodynamic modes that pair with the acoustic modes. In the orthogonal cavity, the 2nd hydrodynamic mode prevails, in which two relatively small vortices travel across the cavity opening simultaneously. In the inclined cavities, however, a single large-scale roll-up vortex corresponding to the 1st hydrodynamic mode is generated owing to the strong Kelvin-Helmholtz instability in the shear layer. More importantly, this vortex spends much of its lifetime growing in size rather than travelling rapidly downstream, resulting in a longer crossing time per cycle that correlates with the 1st acoustic eigenmode frequency ($St=0.276$). Furthermore, aeroacoustic resolvent analysis indicates that inclined cavities amplify acoustic responses more effectively and exhibit weaker source-sink cancellation than the orthogonal cavity. These mechanisms are identified as the primary contributors to the enhanced aeroacoustic response of the inclined cavities. Finally, it is proposed that the ratio of acoustic particle displacement to momentum thickness can be used as a criterion for predicting the onset of deep cavity resonance associated with the distinctive vortex dynamics identified in this paper.
\end{abstract}

\maketitle

\section{Introduction}\label{sec:intro}
Flow-acoustic resonances driven by aeroacoustic instabilities in deep cavity flows produce high-intensity pressure waves at discrete frequencies, leading to detrimental effects such as noise pollution and structural fatigue in various engineering applications. These include safety valves \cite{Coffman1980, Galbally2015}, closed side-branches in gas transport systems \cite{Bruggeman1989, Ziada2010}, turbomachinery \cite{Ziada2002, Aleksentsev2016}, and riverine environments \cite{PerrotMinot2020}. The origin of these resonances lies in the complex interaction between hydrodynamic instabilities and resonant acoustic fields \cite{Bruggeman1989, Peters1993}. When airflow passes over a deep cavity under specific conditions, it can trigger self-sustained oscillations that couple with a depthwise acoustic mode, generating intense aerodynamic noise. In this process, acoustic resonance acts as the primary feedback mechanism, amplifying oscillations and inducing flow-tone lock-ins \cite{Yang2009}. This phenomenon is fundamentally different from oscillations in shallow cavities, which are predominantly governed by the Rossiter feedback mechanism driven by upstream acoustic feedback \cite{Rossiter1964, Rowley2006}. Consequently, a better understanding of the distinct mechanisms driving deep cavity oscillations is therefore essential for mitigating their adverse effects in practical applications.

The aeroacoustics of deep cavity flows have been extensively studied in the scientific literature. Seminal works by \cite{Krishnamurty1955, Plumblee1962, Rossiter1964, East1966} provided experimental evidence that deep cavity flows generate intense acoustic responses near the depthwise acoustic modal frequencies. Rockwell and Naudascher \cite{Rockwell1978} later characterized these phenomena as fluid resonant oscillations driven by the interaction between shear-layer instabilities and depthwise acoustic resonances. In particular, their studies demonstrated that these oscillations first originate from initial shear-layer instabilities near the upstream separation corner. As these instabilities propagate downstream, they interact with the cavity's trailing edge and generate acoustic standing waves. These resonant acoustic fields, in turn, induce velocity perturbations that reinforce shear-layer instabilities, thereby sustaining a closed feedback loop. In deep cavity systems, this loop is particularly pronounced due to the system's inherent susceptibility to minimally radiating depthwise acoustic modes, according to Koch \cite{Koch2005}. As a result, these acoustic resonances can further amplify shear-layer instabilities, giving rise to highly coherent vortex structures frequently observed in deep cavity flows \cite{Tam1978, Tonon2011, Ziada2014, Tinar2014}.

The presence of coherent vortices shows that flow-acoustic resonance in deep cavities is confined to discrete Strouhal-number bands, each linked to a specific hydrodynamic mode of the shear layer. These modes are characterized by the number of convecting vortex structures across the cavity opening that satisfy the requisite streamwise phase criterion \cite{Rowley2002, Tuna2014, ho2021wall}. Among these, flow-acoustic resonances driven by the 1st hydrodynamic mode are well-documented for generating the most intense acoustic responses, predominantly occurring at a Strouhal number of approximately $St\approx 0.4$ \cite{Tonon2011}. In contrast, higher hydrodynamic modes produce weaker resonances at Strouhal numbers exceeding $St\approx 0.8$ \cite{Yang2009, Tonon2011}. Recent experimental investigations on closed side-branches, however, have revealed an additional category of flow-acoustic resonance at a lower Strouhal number, $St\approx 0.27$. This resonance is distinguished by exceptionally strong acoustic responses, surpassing the dynamic pressure of the flow \cite{Peters1993, Ziada1994, Dequand_2003, Bravo2005, Yang2009}. Moreover, within this regime, the resonant acoustic field exerts a significant influence on the coherence and trajectory of vortex structures. Peters \cite{Peters1993} observed that under these conditions, the amplification of instabilities and the intensification of shear-layer oscillations lead to highly nonlinear states, making precise characterization of the fluid resonant mechanism increasingly complex. Consequently, despite substantial empirical evidence, a direct quantitative explanation for the intense, low-frequency resonance at $St\approx 0.27$ remains unresolved.

However, recent advances in modal analysis may offer a promising path toward that explanation. Global linear-stability and receptivity studies have become standard for diagnosing long-term instabilities and uncovering their physical origins across diverse flows \cite{theofilis2000globally, GIANNETTI2007, Bres2008, Theofilis2011, Yamouni2013, Schmid2013, meseguer2014linear, citro2015linear, Marquet2015, liu2016linear, Sun2017}. Direct global and adjoint modes obtained from these methods offer critical insights into structural sensitivity within flow fields \cite{GIANNETTI2007, Schmid2013, Chomaz2005, Marquet2015, soton482472}. Additionally, non-modal approaches, such as resolvent analysis, initially introduced by Trefethen et al. \cite{Trefethen1993} and later extended to turbulent mean flows by McKeon et al. \cite{Mckeon2010} provide a foundational framework for studying energy amplification and the structural response to perturbations over a spectrum of frequencies. These methods have previously been applied to both shallow and deep cavity flows, yielding valuable insights into their underlying dynamics \cite{sun2020resolvent, Liu2021, Liu2021b, Boujo2018}. Nonetheless, the application of resolvent analysis to examine flow-acoustic resonances in deep and inclined cavity flows remains unexplored. Moreover, to our knowledge, this approach has not yet been used to identify the optimal forcing, response, and amplification mechanisms of acoustic perturbations that trigger flow-acoustic resonance in deep cavity configurations.

Presently, most numerical studies have focused primarily on orthogonal geometries \cite{Larchevque2003, thornber2008implicit, sampath2016numerical, chen2017mode, pedergnana2021modeling}. As a result, the mechanisms responsible for noise generation in turbulent flows over deep and inclined cavities, particularly under resonant conditions, remain poorly understood. In this work, we address this gap by employing wall-resolved large-eddy simulations (LES) to investigate the distinct vortex dynamics and noise generation mechanisms in both orthogonal and inclined cavity configurations. The primary objective of this paper is to examine how inclined cavities exhibit aeroacoustic behaviour that differs significantly from that of their orthogonal counterparts. For the subsequent discussions, we consider three inclination angles ($\alpha = 30^{\circ}$, $60^{\circ}$, and $90^{\circ}$) and two free-stream Mach numbers ($M_\infty = 0.2$ and $0.3$). These conditions are selected to highlight the unique aeroacoustic characteristics of deep and inclined cavity flows. It is important to note, however, that the transition mechanisms associated with changes in Mach number or inclination angle are not the central focus of this paper.

This paper is structured and written in the following order. Section \ref{sec:description} outlines the computational setup and methods employed in this study. Sections \ref{sec:pfluc} and \ref{sec:hydrodynamicfield} present a detailed investigation of the acoustic and hydrodynamic fields around the cavity configurations. In Section \ref{sec:inoutputanalysis}, the focus shifts to acoustic amplifications and source-sink cancellations through aeroacoustic resolvent analysis, with particular attention given to the critical role of the ratio of acoustic particle displacement to momentum thickness in defining distinct resonance behaviours. Finally, concluding remarks are provided in Section \ref{sec:conclusion}.

\section{Description of problem and the computational set-up}\label{sec:description}
\begin{figure}
	\centering
	\begin{tikzpicture}
		\node[anchor=south west,inner sep=0] (image) at (0,0) {\includegraphics[width=0.42\textwidth]{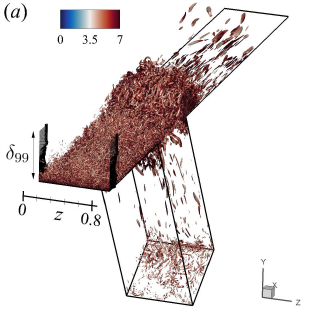}};
		\begin{scope}[x={(image.south east)},y={(image.north west)}]
			\fill [white] (0.0,0.9) rectangle (0.1,1.0);
    		\node[anchor=north west, inner sep=5pt] at (0,1) {(\emph{a})};
		\end{scope}
	\end{tikzpicture}
	\begin{tikzpicture}
	\node[anchor=south west,inner sep=0] (image) at (0,0) {\includegraphics[width=0.42\textwidth]{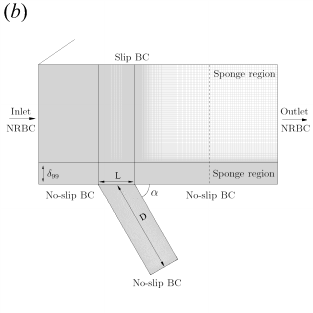}};
	\begin{scope}[x={(image.south east)},y={(image.north west)}]
		\fill [white] (0.0,0.9) rectangle (0.1,1.0);
		\node[anchor=north west, inner sep=5pt] at (0,1) {(\emph{b})};
	\end{scope}
\end{tikzpicture}
	\caption{Visualizations of the current computational domain of the deep and inclined cavity configuration enclosed in a channel. (\emph{a}) Instantaneous non-dimensional $Q$-criterion iso-surfaces ($Q=5$) coloured by non-dimensional vorticity magnitude ($|\omega_{i}|$), unveiling three-dimensional vortices within the turbulent boundary layer. (\emph{b}) A spanwise view of the computational domain used in the current numerical investigation. The cavity length and depth are denoted by $L$ and $D$, respectively.}
	\label{fig:Fig1}
\end{figure}
The present study investigates the cavity section with a length of $L/h=0.608$ and depth of $D/h=1.6$, enclosed in a channel with a height of $2h$, as shown in figure \ref{fig:Fig1}. The Reynolds number based on the cavity opening length, $L = 0.038$ m, is set to $Re_\infty=261,891$ and a freestream Mach number of $M_\infty=0.3$ based on the ambient speed of sound (for air) of $a_\infty=340.2$ m/s and the reference temperature of $T_\infty=288$ K are also considered in this work. The current numerical investigation employs a high-resolution implicit large-eddy simulation (ILES) method based on a wavenumber-optimized discrete filter \cite{Kim2010}. The filter is applied directly to the solution (conservative variables) at every time step and acts as an implicit sub-grid scale (SGS) model that enforces the dissipation of scales smaller than the filter cutoff wavelength. Garmann et al. \cite{Garmann2012} performed an extensive analysis of the ILES technique compared to the traditional implementation of an explicit SGS model and concluded that ILES simulations can correctly capture the flow physics when the grid is subjected to an appropriate resolution.
\subsection{Governing equations and numerical methods}\label{sec:goveqn}
In this work, the full 3-D compressible Navier-Stokes equations (with a source term for sponge layers included) are used, which can be expressed in a conservative form, transformed onto a generalised coordinate system as
\begin{equation}\label{eq:nse}
	\frac{\partial}{\partial t}\left(\frac{\boldsymbol{Q}}{J}\right)+\frac{\partial}{\partial\xi_i}\left(\frac{\boldsymbol{E}_j-Re^{-1}_\infty M_\infty\boldsymbol{F}_j}{J}\frac{\partial\xi_i}{\partial x_j}\right)=-\frac{a_\infty}{L}\frac{\boldsymbol{S}}{J},
\end{equation}
where the indices $i=1,2,3$ and $j=1,2,3$ denote the three dimensions. The vectors of the conservative variables, inviscid and viscous fluxes (that account for losses due to viscous dissipation and thermal conduction), are given by
\begin{equation}\label{eq:Flux}
	\left.
	\begin{gathered}
		\boldsymbol{Q}=[\rho,\rho u,\rho v,\rho w,\rho e_\text{t}]^T,\\
		\boldsymbol{E}_j=[\rho u_j,(\rho uu_j+\delta_{1j}p),(\rho vu_j+\delta_{2j}p),(\rho wu_j+\delta_{3j}p),(\rho e_\text{t}+p)u_j]^T,\\
		\boldsymbol{F}_j=[0, \tau_{1j}, \tau_{2j}, \tau_{3j}, u_i\tau_{ji}+q_j]^T,\\
	\end{gathered}
	\right\}
\end{equation}
with the stress tensor and heat flux vector written as
\begin{equation}
	\tau_{ij}=\mu\left(\frac{\partial u_i}{\partial x_j}+\frac{\partial u_j}{\partial x_i}-\frac{2}{3}\delta_{ij}\frac{\partial u_i}{\partial x_i}\right), \quad q_j=\frac{\mu}{(\gamma-1)Pr}\frac{\partial T}{\partial x_j},
\end{equation}
where $\xi_i=\{\xi,\eta,\zeta\}$ are the generalised coordinates, $x_j=\{x,y,z\}$ are the Cartesian coordinates, $\delta_{ij}$ is the Kronecker delta, $u_j=\{u,v,w\}$, $e_\text{t}=p/[(\gamma-1)\rho]+u_ju_j/2$ and $\gamma=1.4$ for air. The local dynamic viscosity $\mu$ is calculated by using Sutherland's law \cite{White1991}. In the current set-up, $\xi$, $\eta$ and $\zeta$ are aligned in the streamwise, vertical and spanwise directions, respectively. The Jacobian determinant of the coordinate transformation (from Cartesian to the generalised) is given by $J^{-1}=|\partial(x,y,z)/\partial(\xi,\eta,\zeta)|$ \cite{Kim2002AIAA}. The extra source term $\boldsymbol{S}$ on the right-hand side of \eqref{eq:nse} is non-zero within the sponge layer only, which is described in \citet{Kim2010a,Kim2010b}. In this paper, the freestream Mach and Reynolds numbers are defined as $M_\infty=u_\infty/a_\infty$ and $Re_\infty=\rho_\infty u_\infty L/\mu_\infty$ where $a_\infty=\sqrt{\gamma p_\infty/\rho_\infty}$ is the ambient speed of sound and $u_\infty$ is the speed of the freestream mean flow. The governing equations are non-dimensionalised based on the streamwise cavity opening length $L=38$ mm for length scales, the ambient speed of sound $a_\infty$ for velocities, $L/a_\infty$ for time scales and $\rho_\infty a^2_\infty$ for pressure, unless otherwise notified. Temperature, density and dynamic viscosity are normalised by their respective ambient values: $T_\infty$, $\rho_\infty$ and $\mu_\infty$.

The governing equations given above are solved using high-order accurate numerical methods specifically developed for aeroacoustic simulation on structured grids. The flux derivatives in space are calculated based on fourth-order pentadiagonal compact finite difference schemes with seven-point stencils \cite{Kim2007}. Explicit time advancing of the numerical solution is carried out using the classical fourth-order Runge-Kutta scheme with a CFL number of 0.95. Numerical stability is maintained by implementing sixth-order pentadiagonal compact filters for which the cutoff wavenumber (normalized by the grid spacing) is set to $0.85\pi$. In addition to the sponge layers used, characteristics-based non-reflecting boundary conditions (NRBC) based on \cite{Kim2000} are applied at the inflow and outflow boundaries to prevent any outgoing waves from returning to the computational domain. Periodic conditions are used across the spanwise boundary planes unless otherwise stated. Slip (no penetration) and no-slip wall boundary conditions based on \cite{Kim2004} are applied at the top and bottom channel walls, respectively. The top wall boundary is intended to replicate an existing experimental set-up at the University of Southampton. Those who use a different boundary setting on the top boundary, either experimental or computational, will need to take the difference into consideration when they attempt to compare the data.

The computation is parallelized via domain decomposition and message passing interface (MPI) approaches. The compact finite difference schemes and filters used are implicit in space due to the inversion of pentadiagonal matrices involved, which requires a precise and efficient technique for parallelization to avoid numerical artifacts that may appear at the subdomain boundaries. A recent parallelization approach based on quasi-disjoint matrix systems \cite{Kim2013} offering super-linear scalability is used in the present paper.
\subsection{Simulation set-up and discretisation of the problem}\label{sec:setupproblem}
The cavity geometry and the computational domain used in this work comprises $x/L\in[-1.64,4.93]$ in the streamwise direction, $y\in[-2.63,3.29]$ in the vertical direction and $z/L\in[0,0.822]$ in the spanwise direction. The entire computational domain; the inner region (physical domain) where meaningful simulation data are obtained; and, the sponge layer zone is defined as
\begin{equation}\label{eq:domain}
	\left.
	\begin{gathered}
		\mathcal{D}_\infty=\{\xx\,|\,x/L\in[-1.644,4.934],y\in[-2.632,3.289],z/L\in[0,0.822]\},\\
		\mathcal{D}_\text{physical}=\{\xx\,|\,x/L\in[-1.644,3.289],y\in[-2.632,3.289],z/L\in[0,0.822]\},\\
		\mathcal{D}_\text{sponge}=\mathcal{D}_\infty-\mathcal{D}_\text{physical}.
	\end{gathered}
	\right\}
\end{equation}
The physical domain, $\mathcal{D}_\infty$ consists of a deep cavity with an aspect ratio of $D/L=2.632$ enclosed in a straight rectangular channel with a channel half-height of $h/L=1.644$. The channel region is discretised by $960\times 290\times 480$ grid points in streamwise, vertical, and spanwise directions. A total of $240\times 240\times 480$ grid points are used in the streamwise, vertical and spanwise directions, respectively, in the cavity region. The mesh in the wall-normal direction is refined close to the viscous wall $y^{+}\approx 1$ to maintain a sufficiently high level of near-wall grid resolution throughout the viscous wall surfaces. 

{
\renewcommand{\arraystretch}{1.2}
\setlength{\tabcolsep}{14pt} 
  
\begin{table}[!h]
	\vspace{5pt}
	\centering
	\begin{tabular}{l l l l l l}  
    \hline
    $Re_\infty$ & 
    $M_\infty$ & 
    $\alpha$ &
    $\delta^* / L$ &
    $\theta / L$ &
    $H$ \\ 
    \hline
174,594 & 0.2 & $90^{\circ}$ & 0.0434 & 0.0350 & 1.24 \\ 
174,594 & 0.2 & $60^{\circ}$ & 0.0458 & 0.0368 & 1.25 \\ 
174,594 & 0.2 & $30^{\circ}$ & 0.0442 & 0.0355 & 1.24 \\ 
261,891 & 0.3 & $90^{\circ}$ & 0.0379 & 0.0312 & 1.22 \\ 
261,891 & 0.3 & $60^{\circ}$ & 0.0447 & 0.0360 & 1.24 \\ 
261,891 & 0.3 & $30^{\circ}$ & 0.0408 & 0.0332 & 1.23 \\ 
    \hline
	\end{tabular}
	\vspace{5pt}
	\caption{
    The boundary layer information for the current cavity simulations measured at $x=-0.1$.}
	\label{table:Table1}
\end{table}

}

\begin{figure}
	\centering
	\begin{tikzpicture}
	\node[anchor=south west,inner sep=0] (image) at (0,0) {\includegraphics[width=0.33\textwidth]{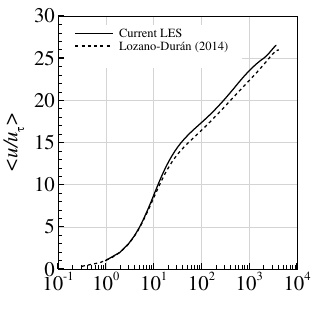}};
	\begin{scope}[x={(image.south east)},y={(image.north west)}]
		\fill [white] (0.0,0.9) rectangle (0.09,1.0);
		\node[anchor=north west, inner sep=0pt] at (0,1) {(\emph{a})};
	\end{scope}
	\end{tikzpicture}
	\begin{tikzpicture}
	\node[anchor=south west,inner sep=0] (image) at (0,0) {\includegraphics[width=0.33\textwidth]{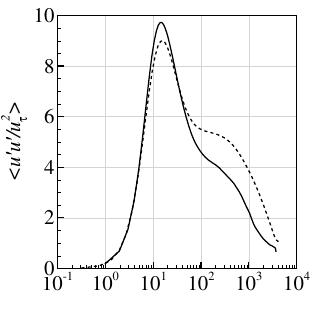}};
	\begin{scope}[x={(image.south east)},y={(image.north west)}]
		\fill [white] (0.0,0.9) rectangle (0.09,1.0);
		\node[anchor=north west, inner sep=0pt] at (0,1) {(\emph{b})};
	\end{scope}
\end{tikzpicture}
	\hspace{5cm}
	\begin{tikzpicture}
	\node[anchor=south west,inner sep=0] (image) at (0,0) {\includegraphics[width=0.33\textwidth]{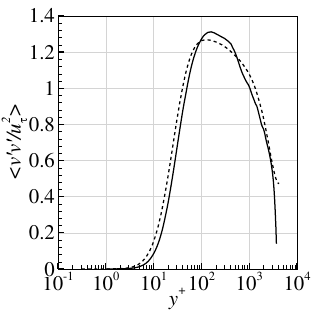}};
	\begin{scope}[x={(image.south east)},y={(image.north west)}]
		\fill [white] (0.0,0.9) rectangle (0.09,1.0);
		\node[anchor=north west, inner sep=0pt] at (0,1) {(\emph{c})};
	\end{scope}
\end{tikzpicture}
	\begin{tikzpicture}
	\node[anchor=south west,inner sep=0] (image) at (0,0) {\includegraphics[width=0.33\textwidth]{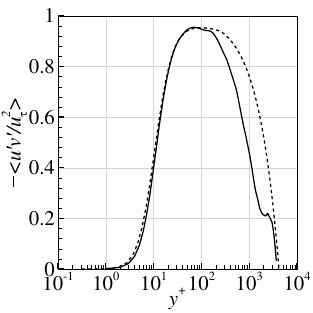}};
	\begin{scope}[x={(image.south east)},y={(image.north west)}]
		\fill [white] (0.0,0.9) rectangle (0.09,1.0);
		\node[anchor=north west, inner sep=0pt] at (0,1) {(\emph{d})};
	\end{scope}
\end{tikzpicture}
	\caption{(\emph{a}) Time-averaged velocity profile of the turbulent boundary layer; and (\emph{b--d}) Reynolds stresses obtained from the current precursor half-channel LES ($Re_{\tau}\approx 3900$), compared with the full-channel DNS ($Re_{\tau}\approx 4200$) by Lozano-Dur{\'a}n and Jim{\'e}nez \cite{Lozano2014}.}
	\label{fig:Fig2}
\end{figure}

The inlet is located at $x/L=-1.664$ upstream of the cavity, where the turbulent inflow data is injected. The outflow boundary is placed at a relatively remote location downstream from the cavity, allowing a sufficient distance for the vortices to dissipate. A precursor simulation is employed to generate the prerequisite turbulent inflow data for the cavity simulation. The precursor simulation domain size ($L_x\times L_y\times L_z$) was set to $4\delta_{99}\times 1\delta_{99}\times 2\delta_{99}$ with $480\times 240\times 480$ grid points in the streamwise, vertical and spanwise directions, respectively. The initial boundary layer thickness, $\delta_{99}$ is determined analytically based on Na and Lu \cite{Na1973}, and the channel flow is initialised with the turbulent mean flow profile according to Spalding \cite{Spalding1961}. In this precursor channel flow simulation, periodic boundary conditions are applied in streamwise and spanwise directions, and a streamwise pressure gradient is applied to maintain the desired mass flow rate. The precursor simulation is completed when the mean flow profile is converged, and the obtained instantaneous flow solutions are injected into the cavity simulation through the inlet plane. Figure \ref{fig:Fig2} shows a close agreement of the time-averaged turbulent velocity profile and the Reynolds stresses between the current half-channel LES and a full-channel DNS by Lozano-Dur{\'a}n and Jim{\'e}nez \cite{Lozano2014}, conducted at $Re_{\tau}\approx$ 3900 and 4200, respectively. The boundary layer data for the current simulations measured at $x=-0.1$ (10\% away from the upstream cavity corner), are listed in Table \ref{table:Table1}.

\subsection{Definition of variables for statistical analysis}\label{sec:statisticalanalysis}
Data processing and analysis are performed upon the completion of the simulation. The main property required in this study is the power spectral density (PSD) function of the pressure fluctuations around the cavity. To facilitate the following discussions, we defined the pressure fluctuations here as
\begin{equation}
	p'(\xx,t)=p(\xx,t)-\overline{p}(\xx),
\end{equation}
where $\overline{p}(\xx)$ is the time-averaged pressure field. Following the definitions used in \citet{Goldstein1976}, the PSD functions of the pressure fluctuations (based on frequency and one-sided) are then calculated by
\begin{equation}
	S_{pp}(\xx,f)=\lim_{T\rightarrow\infty} \frac{\hat{p}(\xx,f,T) \hat{p}^*(\xx,f,T)}{T},
\end{equation}
where $\hat{p}$ is an approximate Fourier transform of $p$, respectively, based on the following definition:
\begin{equation}
	\hat{p}(\xx,f,T)=\int_{-T}^{T}{p'(\xx,t)e^{-2\pi ift}\, \dd t},
\end{equation}
and, `$*$' denotes a complex conjugate. Similarly, the magnitude and the respective phase of the single-sided Fourier transform pressure field are calculated by
\begin{equation}
	\left|p(\xx,f,T)\right|=2 \sqrt{\hat{p}(\xx,f,T) \hat{p}^{*}(\xx,f,T)},
\end{equation}
\begin{equation}
	\Phi_{p}(\xx,f,T)=\arctan \left\{ \frac{ \text{Im}[\hat{p}(\xx,f,T)] }{\text{Re}[ \hat{p}(\xx,f,T)]} \right\}.
\end{equation}
In the above equations, $T$ represents the half-length of the time signals used for the approximate Fourier transform. The same procedures and notation are used for other field quantities later in this paper.

\subsection{Aeroacoustic resolvent analysis}\label{sec:ape}
The Acoustic Perturbation Equations (APEs) proposed by Ewert and Schr\"{o}der \cite{Ewert2003} have been successfully demonstrated as a useful hybrid approach for accurately predicting acoustic propagation within cavity flows by using acoustic sources computed directly from fluid simulations \cite{Aly2012,Arya2021}. In this paper, we employ APEs as a linear operator to explore the dominant input-output characteristics of deep cavity systems based on the time-averaged mean flow states. To achieve this, we incorporate the APE-4 formulation, expressed as
\begin{equation}\label{eq:apecon_time_fulloperator}
	\frac{\partial p'}{\partial t}+\bar{c}^2\nabla \cdot \left(\bar{\rho}\boldsymbol{u}'+\bar{\boldsymbol{u}}\frac{p'}{\bar{c}^2}\right)=\bar{c}^2 q_e,
\end{equation}
\begin{equation}\label{eq:apemom_time_fulloperator}
	\frac{\partial \boldsymbol{u}'}{\partial t} + \nabla\left(\bar{\boldsymbol{u}}\cdot \boldsymbol{u}'\right) + \nabla\left(\frac{p'}{\bar{\rho}}\right)=\boldsymbol{q_m},
\end{equation}
where the noise sources are given by
\begin{equation}\label{eq:apecon_time_fullsource}
	q_{c} = -\nabla \cdot \left(\rho'\boldsymbol{u}'\right)' + \frac{\bar{\rho}}{C_p}\frac{Ds'}{Dt},
\end{equation}
\begin{equation}\label{eq:apemom_time_fullsource}
	\boldsymbol{q_m} = -\left(\boldsymbol{\omega} \times \boldsymbol{u}\right)' + T'\nabla\bar{s} - s'\nabla\bar{T} - \left(\nabla\frac{\boldsymbol{u}'\cdot\boldsymbol{u}'}{2}\right)' + \left(\frac{\nabla\cdot\bar{\boldsymbol{\tau}}}{\rho}\right)'.
\end{equation}
The variables marked with a prime symbol denote fluctuating quantities, whereas those with an overbar represent time-averaged values. Among the source terms, those encapsulating two primed quantities are generally smaller than their counterparts, and consequently, their contribution to the overall sources is considered negligible and thus omitted. In addition, considering the high Reynolds number and relatively low Mach number flow discussed in this paper, the contributions of viscosity and entropy to the sources can be safely omitted. Consequently, the Lamb vector, defined as $(\boldsymbol{\omega} \times \boldsymbol{u})'$, is considered the dominant source term. Applying these simplifications, Eq. (\ref{eq:apecon_time_fulloperator}) and Eq. (\ref{eq:apemom_time_fulloperator}) are rewritten in a compact form, expressed as
\begin{equation}\label{eq:cc}
	\frac{\partial \boldsymbol{q}'}{\partial t}=\boldsymbol{L}(\bar{\boldsymbol{q}})\boldsymbol{q}'+\boldsymbol{f}',
\end{equation}
where $\boldsymbol{L}(\bar{\boldsymbol{q}})$ denotes the linear operator about the mean flow state $\bar{\boldsymbol{q}}=\left[\bar{p},\: \bar{u},\: \bar{v},\: \bar{w} \right]^T$ and $\boldsymbol{f}'$ represents the forcing input comprised of the Lamb vector. Accordingly, a modal perturbation of the form
\begin{equation}\label{eq:lsa_modalperturbation}
	\boldsymbol{q}'(x,y,z,t)=\hat{\boldsymbol{q}}(x,y)\exp{i(\beta z - \omega t)} + \text{complex conjugate},
\end{equation}
is imposed to Eq. (\ref{eq:cc}) to form an input-output dynamics, expressed as
\begin{equation}\label{eq:resolvent0}
	\hat{\boldsymbol{q}}_{\omega} = -[i \omega \boldsymbol{I} + \boldsymbol{L}(\bar{\boldsymbol{q}})]^{-1} \hat{\boldsymbol{f}}_{\omega} = \boldsymbol{R}(\bar{\boldsymbol{q}} ; \omega) \hat{\boldsymbol{f}}_{\omega}.
\end{equation}
Here, the resolvent operator $\boldsymbol{R}(\bar{\boldsymbol{q}} ; \omega)$ relates the input forcing (i.e., Lamb vector), $\hat{\boldsymbol{f}}_{\omega}$, to the output fields as acoustic quantities (i.e., acoustic pressure fields), $\hat{\boldsymbol{q}}_{\omega}$, in the frequency space. The complex eigenvalue is represented by $\omega=\omega_{r} + i \omega_{i}$, with the real part of the eigenvalue, $\omega_{r}$, determining the physical frequency, while its imaginary component determines the radiation loss associated with the acoustic eigenmode ($\omega_i<0$). Furthermore, the acoustic eigenmodes of the cavity systems, which may be influenced by the mean flow field \cite{Aly2012}, can be retrieved by solving the eigenvalue problem presented in the homogeneous form of Eq. (\ref{eq:resolvent0}). Accordingly, the discretized resolvent operator is solved using singular value decomposition to determine the directions spanned by the forcing input and the state output vectors, such as
\begin{equation}\label{eq:resolvent1}
	\boldsymbol{R}(\bar{q};\omega) = \hat{\boldsymbol{U}} \boldsymbol{\Sigma} \hat{\boldsymbol{V}}^{H},
\end{equation}
where $\hat{\boldsymbol{U}} = [\hat{U}_{1},\: \hat{U}_{2},\: \hat{U}_{3},\: \dots]$ and $\hat{\boldsymbol{V}} = [\hat{V}_{1},\: \hat{V}_{2},\: \hat{V}_{3},\: \dots]$ provide the leading optimal sets of responses and the corresponding forcing mode vectors. The amplification gains of the leading optimal sets are determined by the corresponding singular values $\boldsymbol{\Sigma} = \text{diag}(\sigma_{1},\: \sigma_{2},\: \sigma_{3},\: \dots)$, which are arranged in descending order. The superscript $H$ in Eq. (\ref{eq:resolvent1}) indicates the Hermitian transpose operation. In this study, non-penetrating boundary conditions (i.e., zero wall-normal velocity perturbation) are enforced at the wall. Additionally, non-reflecting characteristic boundary conditions introduced by Thompson \cite{Thompson1987, Thompson1990} and damping sponge regions are used in combination to minimize artificial numerical reflections. The approximation of spatial derivatives was achieved using a standard second-order finite difference scheme. Finally, the eigenvalues and eigenvectors of the linear operator were retrieved via the Krylov-Schur algorithm \cite{Stewart2002}. All eigenmodes presented in this paper achieved convergence within a tolerance level of $||\omega\hat{\boldsymbol{Q}}-\boldsymbol{L}\hat{\boldsymbol{Q}}||\leq O(10^{-14})$.

\section{Pressure fluctuations and oscillation frequencies}\label{sec:pfluc}
\begin{figure}[h!]
	\centering
	\begin{tikzpicture}
	\node[anchor=south west,inner sep=0] (image) at (0,0) {\includegraphics[width=0.23\textwidth]{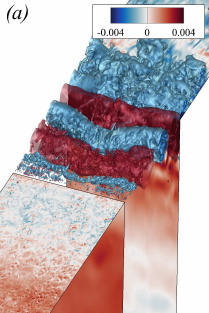}};
	\begin{scope}[x={(image.south east)},y={(image.north west)}]
		\fill [white] (0.0,0.9) rectangle (0.2,1.0);
		\node[anchor=north west, inner sep=1pt] at (0,1) {(\emph{a})};
	\end{scope}
\end{tikzpicture}
	\hspace{0.5cm}
	\begin{tikzpicture}
	\node[anchor=south west,inner sep=0] (image) at (0,0) {\includegraphics[width=0.23\textwidth]{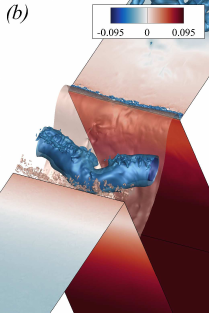}};
	\begin{scope}[x={(image.south east)},y={(image.north west)}]
		\fill [white] (0.0,0.9) rectangle (0.2,1.0);
		\node[anchor=north west, inner sep=1pt] at (0,1) {(\emph{b})};
	\end{scope}
\end{tikzpicture}
	\hspace{0.5cm}
	\begin{tikzpicture}
	\node[anchor=south west,inner sep=0] (image) at (0,0) {\includegraphics[width=0.23\textwidth]{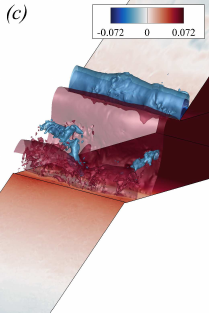}};
	\begin{scope}[x={(image.south east)},y={(image.north west)}]
		\fill [white] (0.0,0.9) rectangle (0.2,1.0);
		\node[anchor=north west, inner sep=1pt] at (0,1) {(\emph{c})};
	\end{scope}
\end{tikzpicture}
	\caption{Large-scale vortical structures are visualized through iso-contours of instantaneous pressure fluctuations, with the flow direction from left to right. Surface contours of wall-pressure fluctuations reveal the prominent acoustic field emanating from deep cavities for inclination angles of (\emph{a}) $\alpha=90^{\circ}$, (\emph{b}) $60^{\circ}$ and (\emph{c}) $30^{\circ}$, respectively.}
	\label{fig:Fig3}
\end{figure}

The self-sustained fluid-resonant oscillation in deep and inclined cavities arises from the interaction between shear-layer fluctuations over the cavity opening and an acoustic mode within the cavity. This interaction amplifies large-scale vortical structures, altering the flow field and producing intense acoustic pressure fluctuations. This process efficiently converts local flow energy into acoustic energy and is illustrated in figure \ref{fig:Fig3}.
\FloatBarrier

\begin{figure}[!h]
	\centering
\begin{tikzpicture}
	\node[anchor=south west,inner sep=0] (image) at (0,0) {\includegraphics[width=0.36\textwidth]{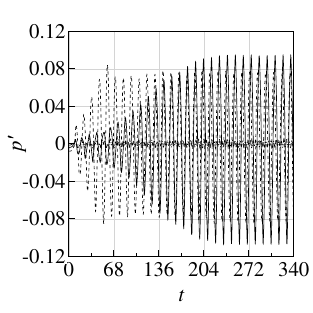}};
	\begin{scope}[x={(image.south east)},y={(image.north west)}]
		\fill [white] (0.0,0.925) rectangle (0.09,1.0);
		\node[anchor=north west, inner sep=1pt] at (0,1) {(\emph{a})};
	\end{scope}
\end{tikzpicture}
\begin{tikzpicture}
	\node[anchor=south west,inner sep=0] (image) at (0,0) {\includegraphics[width=0.36\textwidth]{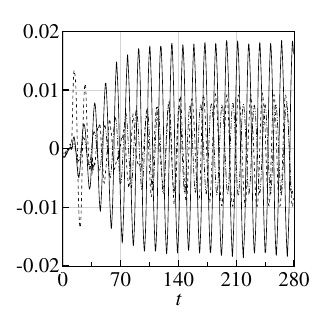}};
	\begin{scope}[x={(image.south east)},y={(image.north west)}]
		\fill [white] (0.0,0.925) rectangle (0.09,1.0);
		\node[anchor=north west, inner sep=1pt] at (0,1) {(\emph{b})};
	\end{scope}
\end{tikzpicture}
	\caption{The spanwise-averaged time signals of pressure fluctuations on the base surface of deep cavities are presented for (\chain) $\alpha=90^{\circ}$, (\full) $60^{\circ}$ and (\dashed) $30^{\circ}$ at free-stream Mach numbers of (\emph{a}) $M_\infty=0.3$ and (\emph{b}) $0.2$, respectively.}
	\label{fig:Fig4}
\end{figure}

This section examines the aeroacoustic behaviour of wall-pressure fluctuations in deep cavities subjected to three distinct inclination angles at two specific Mach numbers, resulting in six simulations. Initially, simulations are conducted using a turbulent inflow dataset at a Mach number of $M_\infty =0.3$ for four million time steps, corresponding to 220 non-dimensional time units. After this period, a steady-periodic state of the wall-pressure signal is achieved at the cavity base for all inclination angles, as shown in Figure~\ref{fig:Fig4}(\emph{a}). Subsequently, the Fourier transform is applied to the pressure time signals over an additional non-dimensional time span of approximately 740 samples (collected every 0.164 time unit) from the computational data, covering a total non-dimensional time of 120. This interval captures approximately ten cycles of the lowest fundamental frequency. The resulting time signals are nearly periodic, and any steady component is eliminated prior to the Fourier transform. Various windowing functions have been tested, and the results exhibit comparable spectrum compositions. The procedures are then repeated with a turbulent inflow dataset at a Mach number of $M_\infty =0.2$, as previously studied by \citet{ho2021wall}. The corresponding wall-pressure signals for each inclination angle are shown in Figure~\ref{fig:Fig4}(\emph{b}). Notably, the time signals from inclined cavities exhibit highly periodic oscillations, highlighting the self-sustaining nature of the oscillation at both Mach numbers.

\begin{figure}[!h]
	\centering
\begin{tikzpicture}
	\node[anchor=south west,inner sep=0] (image) at (0,0) {\includegraphics[width=0.36\textwidth]{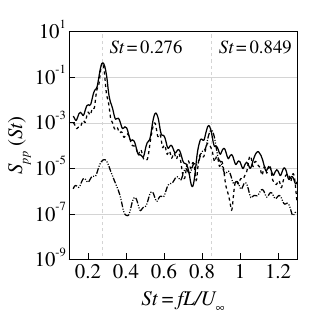}};
	\begin{scope}[x={(image.south east)},y={(image.north west)}]
		\fill [white] (0.0,0.9) rectangle (0.09,1.0);
		\node[anchor=north west, inner sep=5pt] at (0,1) {(\emph{a})};
	\end{scope}
\end{tikzpicture}
\begin{tikzpicture}
	\node[anchor=south west,inner sep=0] (image) at (0,0) {\includegraphics[width=0.36\textwidth]{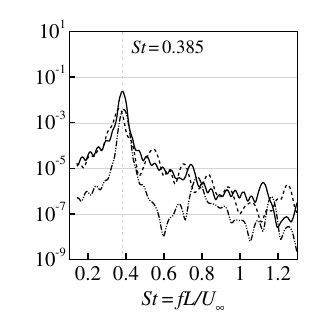}};
	\begin{scope}[x={(image.south east)},y={(image.north west)}]
		\fill [white] (0.0,0.9) rectangle (0.09,1.0);
		\node[anchor=north west, inner sep=5pt] at (0,1) {(\emph{b})};
	\end{scope}
\end{tikzpicture}
	\caption{The power spectral density (PSD) of the spanwise-averaged time signals of wall-pressure fluctuations on the base surface of deep cavities is presented for (\chain) $\alpha=90^{\circ}$, (\full) $60^{\circ}$ and (\dashed) $30^{\circ}$ at free-stream Mach numbers (\emph{a}) $M_\infty=0.3$ and (\emph{b}) $0.2$, respectively.}
	\label{fig:Fig5}
\end{figure}

The power spectra of wall-pressure fluctuations at $M_{\infty}=0.2$ are depicted in figure~\ref{fig:Fig5}(\emph{b}). The figure shows that all three cavity cases ($\alpha=30^{\circ}$, $60^{\circ}$, and $90^{\circ}$) exhibit flow-acoustic resonance closely associated with the fundamental frequency ($St=0.386$). The authors have previously investigated the orthogonal cavity flow characteristics at this fundamental frequency \cite{ho2021wall}, where the critical Mach number for this particular cavity geometry ($D/L=2.632$) and inflow condition ($\theta/L=0.0345$) was estimated to be $M_{\infty}=0.2$. This critical condition was understood to result from a lock-in event between the 1st Rossiter's streamwise feedback and depthwise acoustic resonance modes. Therefore, any deviation in flow speed from this Mach number is expected to produce a sub-optimal flow-acoustic resonance. This assumption is supported by the weaker acoustic response generated from the same orthogonal cavity at $M_{\infty}=0.3$, as shown in figure~\ref{fig:Fig5}(\emph{a}). This weak resonance at $St=0.849$ coincides with the 2nd Rossiter's mode, i.e. $St_n=(n-1/4)/(M_\infty+1/\kappa)$ where $\kappa=0.57$ and $n=2$. However, contrary to previously established expectations, the inclined cavities at this Mach number produce an entirely unexpected result. First, the fundamental peak frequency shifted to a lower value of $St=0.276$ which the Rossiter's model did not predict. It is noteworthy that previous experimental studies on orthogonal deep cavities by \cite{Peters1993,Ziada1994,Dequand_2003,Yang2009} also reported a critical resonance occurring at the similar frequency, which does not conform to existing flow-acoustic resonance theories. Second, and more importantly, the inclined cavities generated a significant increase in the peak amplitude by nearly 30 dB compared to the orthogonal cavity case and by more than 10 dB even compared to the ``optimal'' flow-acoustic resonance at $M_{\infty}=0.2$. The observed shift in the fundamental peak frequency and substantial increase in peak amplitude for inclined cavities at $M_{\infty}=0.3$ suggest novel flow-acoustic interaction mechanisms at play. Consequently, this paper aims to investigate the underlying physical processes responsible for these unexpected and profound results.

\begin{figure}[!h]
	\centering
\begin{tikzpicture}
	\node[anchor=south west,inner sep=0] (image) at (0,0) {\includegraphics[width=0.285\textwidth]{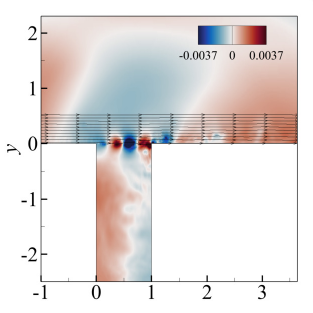}};
	\begin{scope}[x={(image.south east)},y={(image.north west)}]
		\fill [white] (0.0,0.9) rectangle (0.09,1.0);
		\node[anchor=north west, inner sep=0pt] at (0,1) {(\emph{a})};
	\end{scope}
\end{tikzpicture}
\begin{tikzpicture}
	\node[anchor=south west,inner sep=0] (image) at (0,0) {\includegraphics[width=0.285\textwidth]{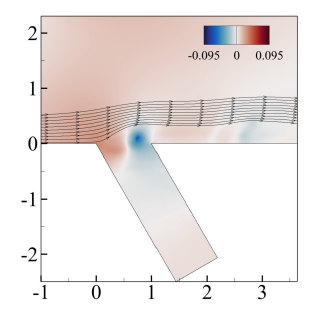}};
	\begin{scope}[x={(image.south east)},y={(image.north west)}]
		\fill [white] (0.0,0.9) rectangle (0.09,1.0);
	\end{scope}
\end{tikzpicture}
\begin{tikzpicture}
	\node[anchor=south west,inner sep=0] (image) at (0,0) {\includegraphics[width=0.285\textwidth]{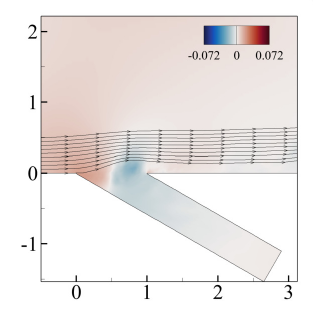}};
	\begin{scope}[x={(image.south east)},y={(image.north west)}]
		\fill [white] (0.0,0.9) rectangle (0.09,1.0);
	\end{scope}
\end{tikzpicture}
    \hspace{5cm}
\begin{tikzpicture}
	\node[anchor=south west,inner sep=0] (image) at (0,0) {\includegraphics[width=0.285\textwidth]{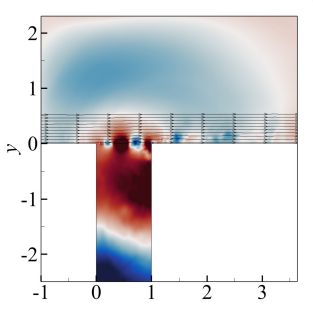}};
	\begin{scope}[x={(image.south east)},y={(image.north west)}]
		\fill [white] (0.0,0.9) rectangle (0.09,1.0);
		\node[anchor=north west, inner sep=0pt] at (0,1) {(\emph{b})};
	\end{scope}
\end{tikzpicture}
\begin{tikzpicture}
	\node[anchor=south west,inner sep=0] (image) at (0,0) {\includegraphics[width=0.285\textwidth]{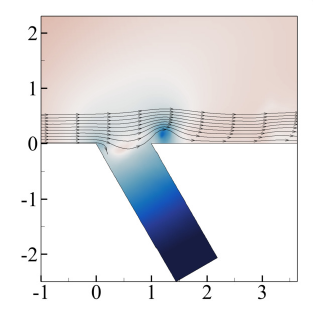}};
	\begin{scope}[x={(image.south east)},y={(image.north west)}]
		\fill [white] (0.0,0.9) rectangle (0.09,1.0);
	\end{scope}
\end{tikzpicture}
\begin{tikzpicture}
	\node[anchor=south west,inner sep=0] (image) at (0,0) {\includegraphics[width=0.285\textwidth]{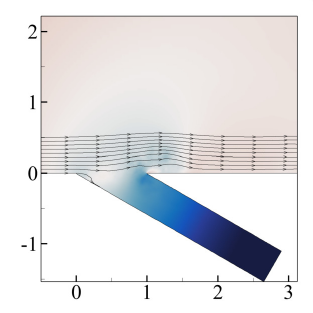}};
	\begin{scope}[x={(image.south east)},y={(image.north west)}]
		\fill [white] (0.0,0.9) rectangle (0.09,1.0);
	\end{scope}
\end{tikzpicture}
    \hspace{5cm}
\begin{tikzpicture}
	\node[anchor=south west,inner sep=0] (image) at (0,0) {\includegraphics[width=0.285\textwidth]{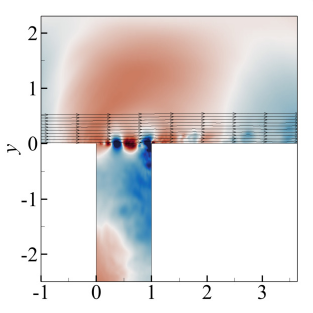}};
	\begin{scope}[x={(image.south east)},y={(image.north west)}]
		\fill [white] (0.0,0.9) rectangle (0.09,1.0);
		\node[anchor=north west, inner sep=0pt] at (0,1) {(\emph{c})};
	\end{scope}
\end{tikzpicture}
\begin{tikzpicture}
	\node[anchor=south west,inner sep=0] (image) at (0,0) {\includegraphics[width=0.285\textwidth]{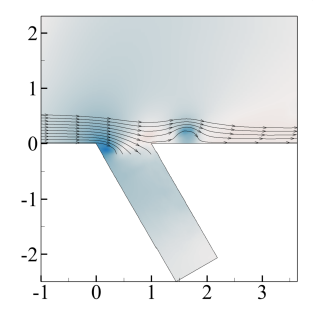}};
	\begin{scope}[x={(image.south east)},y={(image.north west)}]
		\fill [white] (0.0,0.9) rectangle (0.09,1.0);
	\end{scope}
\end{tikzpicture}
\begin{tikzpicture}
	\node[anchor=south west,inner sep=0] (image) at (0,0) {\includegraphics[width=0.285\textwidth]{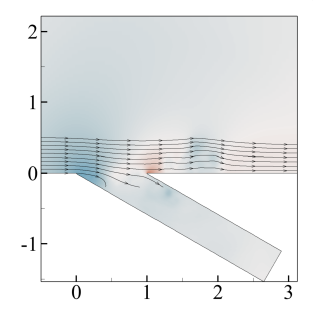}};
	\begin{scope}[x={(image.south east)},y={(image.north west)}]
		\fill [white] (0.0,0.9) rectangle (0.09,1.0);
	\end{scope}
\end{tikzpicture}
    \hspace{5cm}
\begin{tikzpicture}
	\node[anchor=south west,inner sep=0] (image) at (0,0) {\includegraphics[width=0.285\textwidth]{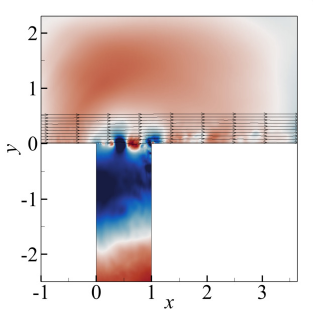}};
	\begin{scope}[x={(image.south east)},y={(image.north west)}]
		\fill [white] (0.0,0.9) rectangle (0.09,1.0);
		\node[anchor=north west, inner sep=0pt] at (0,1) {(\emph{d})};
	\end{scope}
\end{tikzpicture}
\begin{tikzpicture}
	\node[anchor=south west,inner sep=0] (image) at (0,0) {\includegraphics[width=0.285\textwidth]{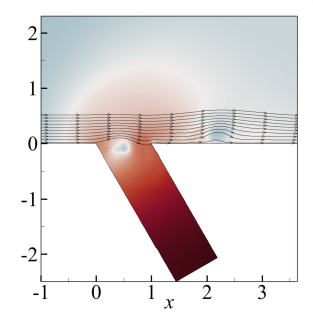}};
	\begin{scope}[x={(image.south east)},y={(image.north west)}]
		\fill [white] (0.0,0.9) rectangle (0.09,1.0);
	\end{scope}
\end{tikzpicture}
\begin{tikzpicture}
	\node[anchor=south west,inner sep=0] (image) at (0,0) {\includegraphics[width=0.285\textwidth]{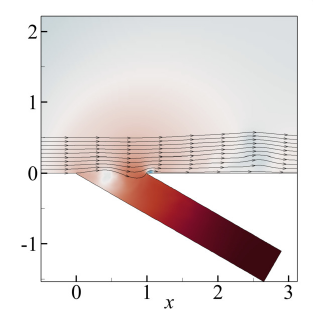}};
	\begin{scope}[x={(image.south east)},y={(image.north west)}]
		\fill [white] (0.0,0.9) rectangle (0.09,1.0);
	\end{scope}
\end{tikzpicture}
    \caption{Snapshots of the spanwise-averaged instantaneous pressure fluctuations $p^\prime$ are shown with superimposed streamlines. The snapshots are taken at time intervals of $T/4$ between successive plots (\emph{a}) to (\emph{d}), where $T$ represents the oscillation cycle period of $\chi$. The first, second, and third columns correspond to deep cavities with $\alpha=90^{\circ}$, $60^{\circ}$ and $30^{\circ}$, respectively. Here, compressive ($p^\prime > 0$) and rarefaction ($p^\prime < 0$) acoustic waves are visualized as red and blue regions within the interior of the cavity.}
    \label{fig:Fig6}
\end{figure}

Figure \ref{fig:Fig6} presents snapshots of the spanwise-averaged instantaneous pressure fluctuations captured at four sequential time intervals, each separated by $T/4$, where $T=1/f_p$ represents the oscillation period corresponding to the tonal frequency identified in the pressure spectra of figure \ref{fig:Fig5}(\emph{a}). These snapshots illustrate the synchronization between shear-layer fluctuations and instantaneous pressure oscillations within the deep and inclined cavities. Here, we define a surface-averaged acoustic pressure at the cavity base to determine the phase of the resonance cycle:
\begin{equation}\label{eq:acousticforceoncf}
	\chi(t) = \frac{1}{A_b}\int_{A_b}^{}{p^\prime(\boldsymbol{x}_b,t)\text{d}A},
\end{equation}
where $x_b$ and $A_b$ denote the Cartesian coordinates on the surface area of the cavity base, respectively. For brevity, the following discussion primarily focuses on the $\alpha=60^{\circ}$ inclined cavity. Figure \ref{fig:Fig6}(\emph{a}) shows the beginning of an oscillation cycle of $\chi$, during which a distinct large-scale vortex is positioned slightly above the cavity opening, as revealed by the low-pressure zone near the downstream corner. At this point, nearly complete destructive interference occurs between the reflected compressive wave (i.e., $p^\prime>0$) and the incident rarefaction acoustic wave (i.e., $p^\prime<0$), resulting in an acoustic pressure equilibrium within the cavity (i.e., $\chi=0$). Immediately thereafter, the rarefaction acoustic wave from the cavity base begins to dominate the cavity and the shear layer moves downward. As the vortex impinges on the downstream corner, it generates additional rarefaction waves that further reduce the acoustic pressure within the cavity toward its minimum value, as depicted in figure \ref{fig:Fig6}($b$). Concurrently, a low‐pressure pocket emerges at the upstream corner, indicating the emergence of the next vortical structure. Subsequently, the main streamlines begin to collide with the downstream corner, inducing flow stagnation and a pressure rise, as shown in figure \ref{fig:Fig6}(\emph{c}). At this instant, another complete destructive interference ($\chi=0$) occurs between the compressive and rarefaction waves. Afterwards, the compressive waves begins to dominate the cavity until $\chi$ reaches its maximum value as depicted in figure \ref{fig:Fig6}(\emph{d}). Throughout this process, the shear layer gradually lifts off the downstream corner as stagnation subsides, allowing the large-scale vortex to convect downstream and be expelled from the cavity, thereby completing a single oscillation cycle and restoring the flow to the configuration shown in figure \ref{fig:Fig6}(\emph{a}).

\begin{figure}[!h]
	\centering
\begin{tikzpicture}
	\node[anchor=south west,inner sep=0] (image) at (0,0) {\includegraphics[width=0.285\textwidth]{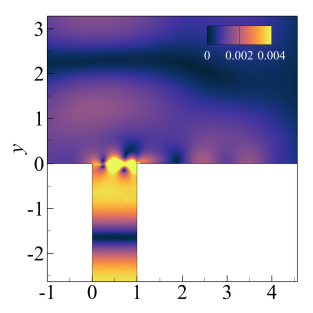}};
	\begin{scope}[x={(image.south east)},y={(image.north west)}]
		\fill [white] (0.0,0.9) rectangle (0.09,1.0);
		\node[anchor=north west, inner sep=1pt] at (0,1) {(\emph{a})};
	\end{scope}
\end{tikzpicture}
\begin{tikzpicture}
	\node[anchor=south west,inner sep=0] (image) at (0,0) {\includegraphics[width=0.285\textwidth]{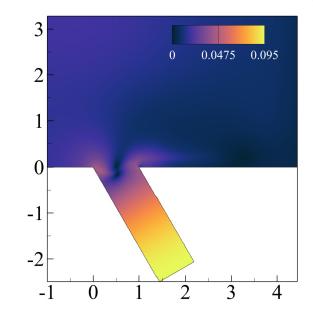}};
	\begin{scope}[x={(image.south east)},y={(image.north west)}]
		\fill [white] (0.0,0.9) rectangle (0.09,1.0);
		\node[anchor=north west, inner sep=1pt] at (0,1) {(\emph{b})};
	\end{scope}
\end{tikzpicture}
\begin{tikzpicture}
	\node[anchor=south west,inner sep=0] (image) at (0,0) {\includegraphics[width=0.285\textwidth]{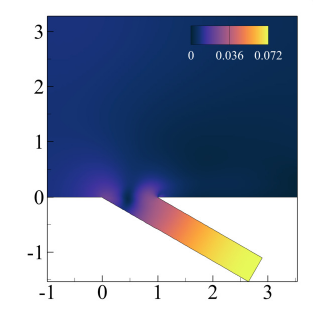}};
	\begin{scope}[x={(image.south east)},y={(image.north west)}]
		\fill [white] (0.0,0.9) rectangle (0.09,1.0);
		\node[anchor=north west, inner sep=1pt] at (0,1) {(\emph{c})};
	\end{scope}
\end{tikzpicture}
	\hspace{5cm}
\begin{tikzpicture}
	\node[anchor=south west,inner sep=0] (image) at (0,0) {\includegraphics[width=0.285\textwidth]{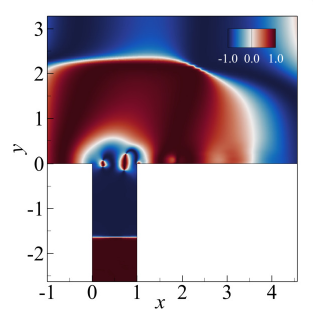}};
	\begin{scope}[x={(image.south east)},y={(image.north west)}]
		\fill [white] (0.0,0.9) rectangle (0.09,1.0);
	\end{scope}
\end{tikzpicture}
\begin{tikzpicture}
	\node[anchor=south west,inner sep=0] (image) at (0,0) {\includegraphics[width=0.285\textwidth]{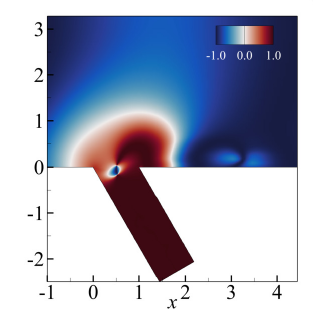}};
	\begin{scope}[x={(image.south east)},y={(image.north west)}]
		\fill [white] (0.0,0.9) rectangle (0.09,1.0);
	\end{scope}
\end{tikzpicture}
\begin{tikzpicture}
	\node[anchor=south west,inner sep=0] (image) at (0,0) {\includegraphics[width=0.285\textwidth]{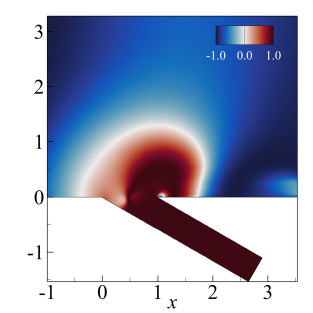}};
	\begin{scope}[x={(image.south east)},y={(image.north west)}]
		\fill [white] (0.0,0.9) rectangle (0.09,1.0);
	\end{scope}
\end{tikzpicture}
	\caption{The spatial distribution of Fourier-transformed pressure fluctuations at the tonal frequency is shown for deep cavities with (\emph{a}) $\alpha=90^{\circ}$, (\emph{b}) $60^{\circ}$ and (\emph{c}) $30^{\circ}$, respectively. The top row of the contour plots represents the magnitude $|p'|$, while the bottom row of the contour plots represents the cosine of the phase $\cos[\Phi_{p'}(\xx,f)-\Phi_{\chi}(\xx,f)]$. Here, $\Phi_{\chi}(\xx,f)$ denotes the phase of the Fourier transform of $\chi$, as defined in Eq. (\ref{eq:acousticforceoncf}).}
	\label{fig:Fig7}
\end{figure}
The discussions above implicitly identified two main types of pressure fluctuations, namely the local hydrodynamic fluctuations near the cavity opening and acoustic fluctuations surrounding the cavity. The distinction between these two components can be further clarified through the magnitude and phase distributions of the Fourier-transformed pressure fluctuations. Accordingly, figure \ref{fig:Fig7} shows that the pressure field inside the cavity appears to be primarily stationary (i.e., constant phase), with a maximum magnitude (i.e., a pressure node) consistently located at the cavity base across all inclinations. This evidence suggests that the pressure field inside the cavity is predominantly acoustic in nature and contributed by the depthwise acoustic resonances. Furthermore, these resonances are highly localized within the inclined cavities and resemble the ``nearly trapped acoustic mode'', that exhibits minimal radiation losses, according to Koch \cite{Koch2005}.

To facilitate subsequent investigations, we decompose the pressure fluctuations around the cavity into their hydrodynamic and acoustic components using momentum potential theory (MPT) developed by Doak \cite{Doak1989}. Essentially, Doak's MPT separates the momentum density, $\rho\boldsymbol{u}$, into rotational and irrotational components through a Helmholtz decomposition. The Helmholtz decomposition of $\rho\boldsymbol{u}$ may be written as
\begin{equation}\label{eq:doak0}
	\rho\boldsymbol{u}=\boldsymbol{B}-\nabla\psi, \quad \nabla \boldsymbol{\cdot} \boldsymbol{B} = 0,
\end{equation}
where $\boldsymbol{B}$ and $\nabla\psi$ are the solenoidal and irrotational components of $\rho\boldsymbol{u}$, respectively. Substituting Eq. (\ref{eq:doak0}) into the continuity equation yields a Poisson equation for the irrotational component, with a source term dependent on density fluctuation,
\begin{equation}\label{eq:doak1}
	\nabla ^2 \psi = \frac{\partial \rho}{\partial t}.
\end{equation}
For a single phase continuum fluid, $\psi$ is separated into acoustic component (irrotational and isentropic, denoted $\psi_{A}$) and entropic component (irrotational and isobaric, $\psi_{E}$) components, governed by the exact equations
\begin{equation}\label{eq:doak2}
	\psi=\psi_{A}+\psi_{E}, \quad \nabla^2 \psi_{A}=\frac{1}{c^2}\frac{\partial \rho}{\partial t}, \quad \nabla^2\psi_{E}=\frac{\partial \rho}{\partial E}\frac{\partial E}{\partial t}.
\end{equation}
Considering the low Mach number in this study, the entropy (thermal) contribution is assumed to be relatively small compared to the acoustic contribution, and therefore $\psi_E$ is not included in the subsequent calculation. Then, the momentum equation in terms of the hydrodynamic and acoustic components is obtained by substituting Eq. (\ref{eq:doak0}) into the momentum equation, expressed as
\begin{equation}\label{eq:doak3}
	\frac{\partial}{\partial t}(\boldsymbol{B}-\nabla\psi) + \nabla \boldsymbol{\cdot} \left[\frac{(\boldsymbol{B}-\nabla\psi)(\boldsymbol{B}-\nabla\psi)}{\rho}-\tau_{ij}\right] + \nabla p =0.
\end{equation}
By taking the divergence of Eq. (\ref{eq:doak3}), the Poisson equation for the hydrodynamic pressure fluctuation, $p'_{H}$
\begin{equation}\label{eq:doak4}
	\nabla^2 p'_{H} = S_{H} +\tilde{S}_H,
\end{equation}
and the Poisson equation for the acoustic pressure fluctuation, $p'_{A}$
\begin{equation}\label{eq:doak5}
	\nabla^2 p'_{A} = S_{A} +\tilde{S}_A,
\end{equation}
are derived. Accordingly, the hydrodynamic and acoustic pressure fluctuations are obtained by solving the Poisson equations in Eq. (\ref{eq:doak4}) and Eq. (\ref{eq:doak5}), respectively. The numerical implementation is described extensively in \cite{Unnikrishnan2016, ho2021wall} and the evaluations of the linear ($S_{H}$ and $S_{A}$) and the non-linear source terms ($\tilde{S}_H$ and $\tilde{S}_A$) are detailed in \cite{Unnikrishnan2020}, which are not repeated here for brevity.

\begin{figure}[!h]
	\centering
\begin{tikzpicture}
	\node[anchor=south west,inner sep=0] (image) at (0,0) {\includegraphics[width=0.285\textwidth]{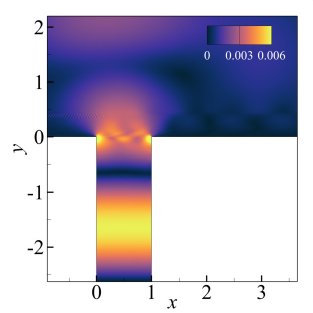}};
	\begin{scope}[x={(image.south east)},y={(image.north west)}]
		\fill [white] (0.0,0.9) rectangle (0.09,1.0);
		\node[anchor=north west, inner sep=1pt] at (0,1) {(\emph{a})};
	\end{scope}
\end{tikzpicture}
\begin{tikzpicture}
	\node[anchor=south west,inner sep=0] (image) at (0,0) {\includegraphics[width=0.285\textwidth]{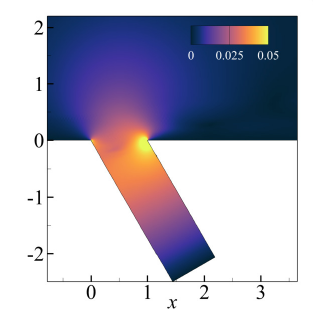}};
	\begin{scope}[x={(image.south east)},y={(image.north west)}]
		\fill [white] (0.0,0.9) rectangle (0.09,1.0);
		\node[anchor=north west, inner sep=1pt] at (0,1) {(\emph{b})};
	\end{scope}
\end{tikzpicture}
\begin{tikzpicture}
	\node[anchor=south west,inner sep=0] (image) at (0,0) {\includegraphics[width=0.285\textwidth]{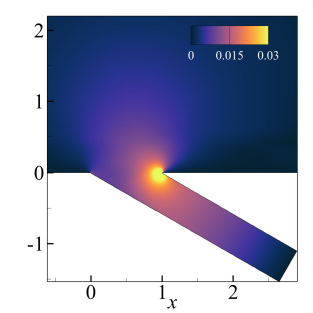}};
	\begin{scope}[x={(image.south east)},y={(image.north west)}]
		\fill [white] (0.0,0.9) rectangle (0.09,1.0);
		\node[anchor=north west, inner sep=1pt] at (0,1) {(\emph{c})};
	\end{scope}
\end{tikzpicture}
    \caption{The contour plots illustrate the spatial distribution of the decomposed Fourier-transformed pressure gradient magnitude associated with the acoustic component $|\partial p_{A}/\partial y|$ at the tonal frequency for (\emph{a}) $\alpha=90^{\circ}$, (\emph{b}) $60^{\circ}$ and (\emph{c}) $30^{\circ}$.}
	\label{fig:Fig8}
\end{figure}
Figure \ref{fig:Fig8} reveals notable differences in the spatial distribution of pressure gradients for the acoustic components across orthogonal and inclined cavities. In particular, the acoustic pressure gradient tends to concentrate more intensely near the downstream corner in inclined cavities, in contrast to the symmetric distribution observed in the orthogonal cavity. This discrepancy in the spatial distribution of the acoustic pressure gradient is important for elucidating the noise generation process, which will be further discussed in Section \ref{sec:inoutputanalysis}. Additionally, the difference in magnitude of the acoustic pressure gradient indicates that the resonant acoustic fields in inclined cavities may produce higher acoustic particle velocities compared to those in the orthogonal cavity. To quantify these observations, the induced acoustic particle velocity along the cavity opening region is approximated as being proportional to the acoustic pressure gradient, using the isentropic Euler equations \cite{Rienstra2015}, as expressed by
\begin{equation}\centering
	\frac{d v_{a}}{dt}=-\frac{1}{\rho}\frac{\partial p_{a}}{\partial y},
\end{equation}
where $\tilde{v}_{a}$ represents the estimated acoustic particle velocity and $p_{a}$ is the decomposed acoustic pressure field. Then, by considering a modal fluctuation of the acoustic pressure and spatially averaging the acoustic particle velocity across the cavity opening, we obtain an averaged acoustic particle velocity that oscillates across the cavity opening, as given by
\begin{equation}\centering
	\overline{v}_{a}=\frac{1}{L}\int_{x=0}^{x=L}{\frac{1}{2\pi f}\frac{\partial p_{a}}{\partial y} \: \text{d}x}.
\end{equation}
At tonal frequencies, cavities with inclinations of ($\alpha=$ 90$^{\circ}$, 60$^{\circ}$, and 30$^{\circ}$) exhibit an average acoustic particle velocity across the cavity opening of approximately ($|\overline{v}_{a}|/u_\infty \approx 9.3\times 10^{-3}$, $2.5\times 10^{-1}$, and $1.1\times 10^{-1}$), respectively. For inclined cavities, these amplitude levels are traditionally categorized as ``high pulsation levels'', according to Bruggeman \cite{Bruggemanthesis,Bruggeman1989}. Previous studies by Peters \cite{Peters1993} have demonstrated that under such extreme conditions, the resonant field can significantly alter the vortex trajectory, causing the vortex to enter and exit the cavity rather than following the parallel path of an unperturbed shear layer. These intense flow dynamics align with the temporal evolution of instantaneous pressure fluctuations and shear-layer oscillations observed in figure \ref{fig:Fig6}. In contrast, the acoustic particle velocity magnitude in the orthogonal cavity corresponds to ``low pulsation levels'', reflecting the subdued shear-layer oscillation. This behavior will be further discussed and visualized later in figure \ref{fig:Fig13}.

\begin{figure}[!h]
	\centering
\begin{tikzpicture}
	\node[anchor=south west,inner sep=0] (image) at (0,0) {\includegraphics[width=0.36\textwidth]{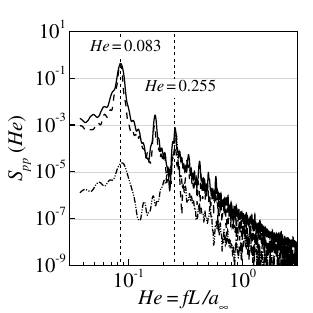}};
	\begin{scope}[x={(image.south east)},y={(image.north west)}]
		\fill [white] (0.0,0.9) rectangle (0.09,1.0);
		\node[anchor=north west, inner sep=3pt] at (0,1) {(\emph{a})};
	\end{scope}
\end{tikzpicture}
\begin{tikzpicture}
	\node[anchor=south west,inner sep=0] (image) at (0,0) {\includegraphics[width=0.36\textwidth]{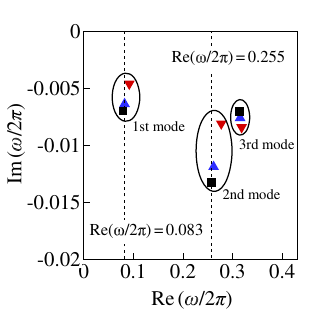}};
	\begin{scope}[x={(image.south east)},y={(image.north west)}]
		\fill [white] (0.0,0.9) rectangle (0.09,1.0);
		\node[anchor=north west, inner sep=3pt] at (0,1) {(\emph{b})};
	\end{scope}
\end{tikzpicture}
	\caption{Panel (\emph{a}) presents the power spectral density (PSD) of the spanwise-averaged wall-pressure fluctuations for (\chain) $\alpha=90^{\circ}$, (\full) $60^{\circ}$ and (\dashed) $30^{\circ}$ at $M_\infty=0.3$, with frequency expressed as the Helmholtz number ($He=fL/a_\infty$). The tonal frequencies from the LES, indicated by the vertical dashed lines (\dashed) at $He=0.083$ and $He=0.255$, are compared in panel (\emph{b}) with the first three least-damped depthwise acoustic modes obtained from the modal analysis of APEs for cavities with (\protect\filledbox) $\alpha=90^{\circ}$, (\protect\filledtriangleblue) $60^{\circ}$ and (\protect\filledgradientred) $30^{\circ}$. Here, the orthogonal cavity produced a resonance with a 2nd depthwise acoustic mode while the inclined cavities with a 1st mode. The 1st depthwise acoustic mode exhibits less radiation losses with more acoustic energy contained within the cavity to excite the shear layer. Furthermore, the frequencies of the depthwise acoustic modes closely match the first classical acoustic quarter-wave (e.g., $He=0.095$) and the third acoustic quarter-wave (e.g., $He=0.285$) of a closed tube. In experimental studies, any slight discrepancies in frequency from this classical prediction are typically minimized by applying end corrections, see Yang et al \cite{Yang2009}, for example.}
	\label{fig:Fig9}
\end{figure}
\begin{figure}[!h]
	\centering
\begin{tikzpicture}
	\node[anchor=south west,inner sep=0] (image) at (0,0) {\includegraphics[width=0.285\textwidth]{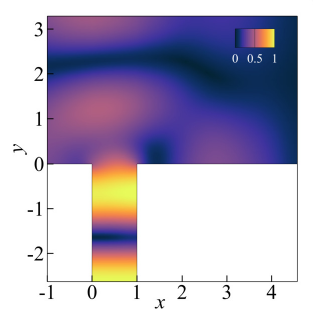}};
	\begin{scope}[x={(image.south east)},y={(image.north west)}]
		\fill [white] (0.0,0.9) rectangle (0.09,1.0);
		\node[anchor=north west, inner sep=0pt] at (0,1) {(\emph{a})};
	\end{scope}
\end{tikzpicture}
\begin{tikzpicture}
	\node[anchor=south west,inner sep=0] (image) at (0,0) {\includegraphics[width=0.285\textwidth]{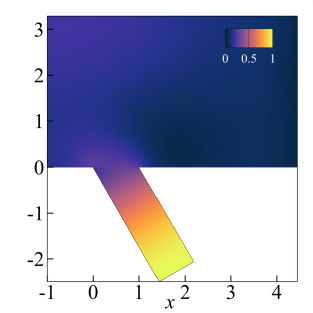}};
	\begin{scope}[x={(image.south east)},y={(image.north west)}]
		\fill [white] (0.0,0.9) rectangle (0.09,1.0);
		\node[anchor=north west, inner sep=0pt] at (0,1) {(\emph{b})};
	\end{scope}
\end{tikzpicture}
\begin{tikzpicture}
	\node[anchor=south west,inner sep=0] (image) at (0,0) {\includegraphics[width=0.285\textwidth]{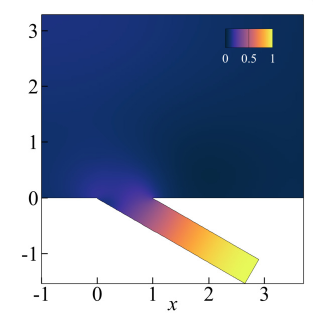}};
	\begin{scope}[x={(image.south east)},y={(image.north west)}]
		\fill [white] (0.0,0.9) rectangle (0.09,1.0);
		\node[anchor=north west, inner sep=0pt] at (0,1) {(\emph{c})};
	\end{scope}
\end{tikzpicture}
	\caption{The contour plots depict the spatial distribution of the pressure magnitude for the eigenmode at frequency nearest to the tonal frequency observed in cavities with (\emph{a}) $\alpha=90^{\circ}$, (\emph{b}) $60^{\circ}$ and (\emph{c}) $30^{\circ}$. For fair comparisons, the pressure magnitude is normalised such that it is unity at the base of the cavity.}
	\label{fig:Fig10}
\end{figure}
Figure \ref{fig:Fig9}(\emph{a}) presents a comparison of pressure spectra for all cavity oscillations at $M_\infty=0.3$, similar to that shown in figure \ref{fig:Fig5}(\emph{a}), with frequencies here expressed in Helmholtz numbers, $He=fL/a_\infty$, to enable direct comparison with the first three least-damped acoustic modes (eigenmodes) obtained from the modal analysis of APEs. These acoustic eigenmodes are characterized by complex resonance frequencies, with the real part indicating the physical resonance frequencies and the imaginary part measuring the radiation losses (i.e., damping) of the resonances, as outlined in Subsection \ref{sec:ape}. Here, the damping levels may provide useful indications of the relative amplitudes of resonance between different modes when an identical acoustic input is imposed. It worth mentioning here that the eigenmode analysis is purely based on acoustic perturbation with no flow fluctuation (feedback) involved. The eigenmodes analysis indicates possible frequencies (or around them) at which an acoustic resonance may occur. The damping levels may be indicative of the relative amplitudes of resonance between different modes when an identical acoustic input is imposed without feedback from the flow.

Figure \ref{fig:Fig9}(\emph{b}) shows that the tonal frequencies of the cavity oscillations at $M_\infty=0.3$ all reside close to their respective acoustic eigenmodes, highlighting the important role of acoustic resonances in supporting the oscillation frequency in deep and inclined cavities. When comparing the orthogonal and inclined cavity cases, a 1st depthwise acoustic mode is excited in the inclined cavities which have lower radiation losses than the 2nd mode excited in the orthogonal cavity. As discussed by Koch \cite{Koch2005}, acoustic modes with higher radiation losses (i.e., a more negative imaginary part of the complex frequency) radiate and dissipate more energy into the surroundings and less energy contained within the cavity. This is illustrated in figure \ref{fig:Fig10}, where the inclined cavities (1st mode) show much less leakage of acoustic energy into the far field than the orthogonal cavity (2nd mode). The elevated acoustic energy retained within the inclined cavities may allow them to excite the shear layer with greater magnitude. Here, it is worth highlighting that the selection of oscillation frequency in deep and inclined cavities is governed by multiple factors, with disparities in radiation losses of depthwise acoustic modes being only one contributing aspect. In the following sections, we examine how the vortex dynamics (resulting from shear-layer perturbations) and the acoustic response of the cavity influence frequency selection, particularly at $St= 0.276$ in inclined cavities.

\section{Hydrodynamic fields}\label{sec:hydrodynamicfield}
In this section, we discuss the hydrodynamic fields near the cavity opening in detail. As mentioned, the location of the coherent vortical structure is crucial to the acoustic emission process. Therefore, an accurate description of the position and path traveled by the vortical structure, which is a function of time, is essential for this investigation. Generally, the location of the vortical structure can be approximated using the pressure minima technique, as shown in Section \ref{sec:pfluc}. However, it is challenging to justify an accurate quantification of the hydrodynamic mode based on the number of discrete low-pressure spots \cite{ho2021wall}. To overcome this limitation, the location of the vortical structure is identified using the equivalent $Q$-criterion \cite{Bradshaw1981}, which is given by
\begin{equation}\label{eq:div2p}
	Q= \epsilon_{ij}\epsilon_{ji} - \frac{1}{2} \omega_{i}^2 \approx -\nabla^{2} \tilde{p_{H}}/\rho_{\infty},
\end{equation}
where $\epsilon_{ij}=\frac{1}{2} ( \partial u_i/ \partial x_j + \partial u_j / \partial x_i)$ represents the rate of strain, $\omega_{i}$ denotes the vorticity of the velocity field, and $\nabla^{2} \tilde{p_{H}}$ is the Laplacian of the hydrodynamic pressure field. This formulation offers two distinct advantages: first, Eq. (\ref{eq:div2p}) establishes a direct link between the velocity gradient field and the hydrodynamic pressure field to accurately pinpoint the location of the vortex. Second, the positive and negative values of the $Q$-criterion provide valuable insights into the strain rate and vorticity of the velocity fields, which are essential for understanding subsequent noise generation mechanisms \cite{Naguib2004}.

\begin{figure}[!h]
	\centering
\begin{tikzpicture}
	\node[anchor=south west,inner sep=0] (image) at (0,0) {\includegraphics[width=0.29\textwidth]{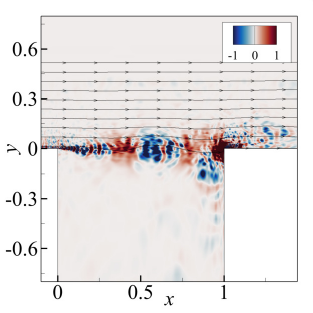}};
	\begin{scope}[x={(image.south east)},y={(image.north west)}]
		\fill [white] (0.0,0.9) rectangle (0.09,1.0);
		\node[anchor=north west, inner sep=1pt] at (0,1) {(\emph{a})};
	\end{scope}
\end{tikzpicture}
\begin{tikzpicture}
	\node[anchor=south west,inner sep=0] (image) at (0,0) {\includegraphics[width=0.29\textwidth]{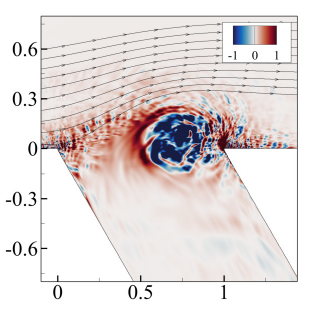}};
	\begin{scope}[x={(image.south east)},y={(image.north west)}]
		\fill [white] (0.0,0.9) rectangle (0.09,1.0);
	\end{scope}
\end{tikzpicture}
\begin{tikzpicture}
	\node[anchor=south west,inner sep=0] (image) at (0,0) {\includegraphics[width=0.29\textwidth]{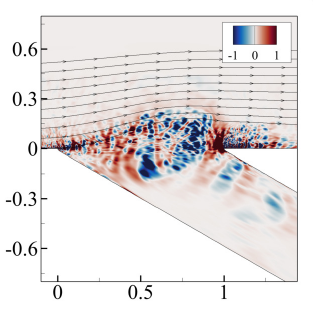}};
	\begin{scope}[x={(image.south east)},y={(image.north west)}]
		\fill [white] (0.0,0.9) rectangle (0.09,1.0);
	\end{scope}
\end{tikzpicture}
    \hspace{5cm}
\begin{tikzpicture}
	\node[anchor=south west,inner sep=0] (image) at (0,0) {\includegraphics[width=0.29\textwidth]{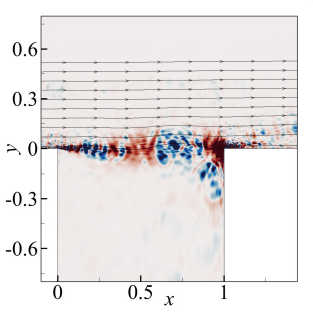}};
	\begin{scope}[x={(image.south east)},y={(image.north west)}]
		\fill [white] (0.0,0.9) rectangle (0.09,1.0);
		\node[anchor=north west, inner sep=1pt] at (0,1) {(\emph{b})};
	\end{scope}
\end{tikzpicture}
\begin{tikzpicture}
	\node[anchor=south west,inner sep=0] (image) at (0,0) {\includegraphics[width=0.29\textwidth]{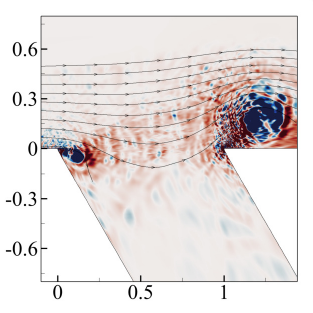}};
	\begin{scope}[x={(image.south east)},y={(image.north west)}]
		\fill [white] (0.0,0.9) rectangle (0.09,1.0);
	\end{scope}
\end{tikzpicture}
\begin{tikzpicture}
	\node[anchor=south west,inner sep=0] (image) at (0,0) {\includegraphics[width=0.29\textwidth]{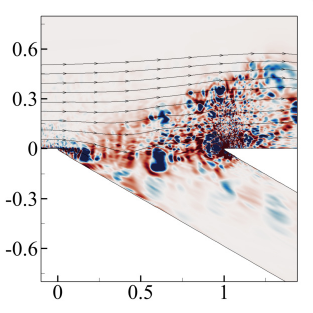}};
	\begin{scope}[x={(image.south east)},y={(image.north west)}]
		\fill [white] (0.0,0.9) rectangle (0.09,1.0);
	\end{scope}
\end{tikzpicture}
    \hspace{5cm}
\begin{tikzpicture}
	\node[anchor=south west,inner sep=0] (image) at (0,0) {\includegraphics[width=0.29\textwidth]{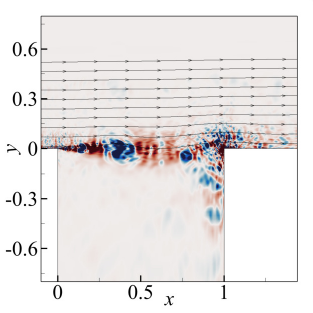}};
	\begin{scope}[x={(image.south east)},y={(image.north west)}]
		\fill [white] (0.0,0.9) rectangle (0.09,1.0);
		\node[anchor=north west, inner sep=1pt] at (0,1) {(\emph{c})};
	\end{scope}
\end{tikzpicture}
\begin{tikzpicture}
	\node[anchor=south west,inner sep=0] (image) at (0,0) {\includegraphics[width=0.29\textwidth]{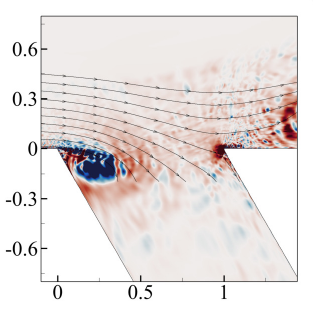}};
	\begin{scope}[x={(image.south east)},y={(image.north west)}]
		\fill [white] (0.0,0.9) rectangle (0.09,1.0);
	\end{scope}
\end{tikzpicture}
\begin{tikzpicture}
	\node[anchor=south west,inner sep=0] (image) at (0,0) {\includegraphics[width=0.29\textwidth]{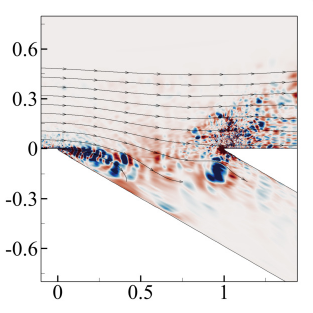}};
	\begin{scope}[x={(image.south east)},y={(image.north west)}]
		\fill [white] (0.0,0.9) rectangle (0.09,1.0);
	\end{scope}
\end{tikzpicture}
    \hspace{5cm}
\begin{tikzpicture}
	\node[anchor=south west,inner sep=0] (image) at (0,0) {\includegraphics[width=0.29\textwidth]{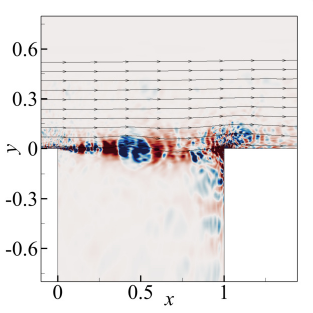}};
	\begin{scope}[x={(image.south east)},y={(image.north west)}]
		\fill [white] (0.0,0.9) rectangle (0.09,1.0);
		\node[anchor=north west, inner sep=1pt] at (0,1) {(\emph{d})};
	\end{scope}
\end{tikzpicture}
\begin{tikzpicture}
	\node[anchor=south west,inner sep=0] (image) at (0,0) {\includegraphics[width=0.29\textwidth]{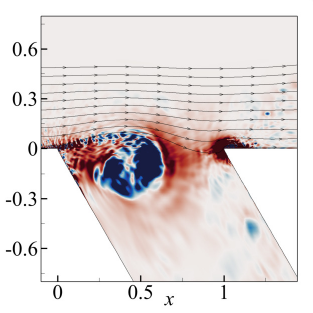}};
	\begin{scope}[x={(image.south east)},y={(image.north west)}]
		\fill [white] (0.0,0.9) rectangle (0.09,1.0);
	\end{scope}
\end{tikzpicture}
\begin{tikzpicture}
	\node[anchor=south west,inner sep=0] (image) at (0,0) {\includegraphics[width=0.29\textwidth]{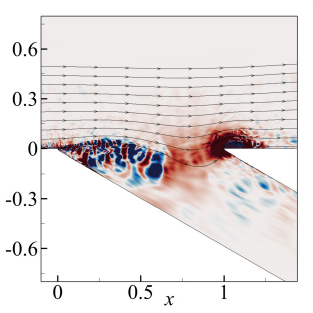}};
	\begin{scope}[x={(image.south east)},y={(image.north west)}]
		\fill [white] (0.0,0.9) rectangle (0.09,1.0);
	\end{scope}
\end{tikzpicture}
    \caption{Snapshots of the spanwise-averaged instantaneous $Q$-criterion with superimposed streamlines to signify the shear-layer undulation across the cavity opening with a time interval of $T$/4 between two successive plots from (\emph{a}) to (\emph{d}), where $T$ is the period of the oscillation cycle of $\chi$. The first, second and third columns correspond to $\alpha=90^{\circ}$, $60^{\circ}$ and $30^{\circ}$, respectively. For the corresponding hydrodynamic pressure fields, refer to figure \ref{fig:Fig12}.}
	\label{fig:Fig11}
\end{figure}
\begin{figure}[!h]
	\centering
\begin{tikzpicture}
	\node[anchor=south west,inner sep=0] (image) at (0,0) {\includegraphics[width=0.29\textwidth]{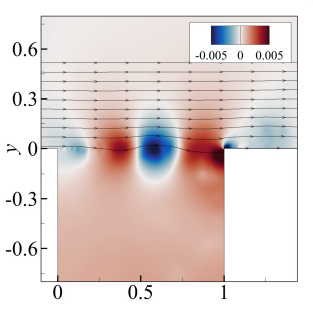}};
	\begin{scope}[x={(image.south east)},y={(image.north west)}]
		\fill [white] (0.0,0.9) rectangle (0.09,1.0);
		\node[anchor=north west, inner sep=1pt] at (0,1) {(\emph{a})};
	\end{scope}
\end{tikzpicture}
\begin{tikzpicture}
	\node[anchor=south west,inner sep=0] (image) at (0,0) {\includegraphics[width=0.29\textwidth]{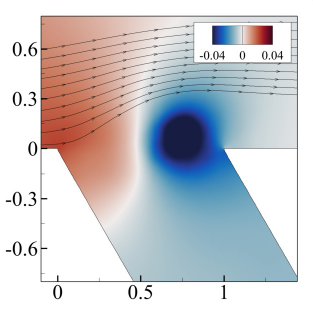}};
	\begin{scope}[x={(image.south east)},y={(image.north west)}]
		\fill [white] (0.0,0.9) rectangle (0.09,1.0);
	\end{scope}
\end{tikzpicture}
\begin{tikzpicture}
	\node[anchor=south west,inner sep=0] (image) at (0,0) {\includegraphics[width=0.29\textwidth]{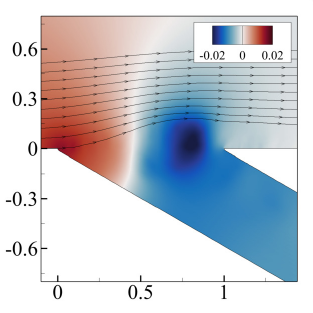}};
	\begin{scope}[x={(image.south east)},y={(image.north west)}]
		\fill [white] (0.0,0.9) rectangle (0.09,1.0);
	\end{scope}
\end{tikzpicture}
    \hspace{5cm}
\begin{tikzpicture}
	\node[anchor=south west,inner sep=0] (image) at (0,0) {\includegraphics[width=0.29\textwidth]{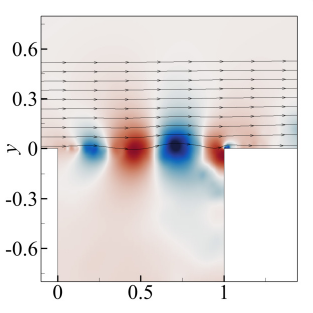}};
	\begin{scope}[x={(image.south east)},y={(image.north west)}]
		\fill [white] (0.0,0.9) rectangle (0.09,1.0);
		\node[anchor=north west, inner sep=1pt] at (0,1) {(\emph{b})};
	\end{scope}
\end{tikzpicture}
\begin{tikzpicture}
	\node[anchor=south west,inner sep=0] (image) at (0,0) {\includegraphics[width=0.29\textwidth]{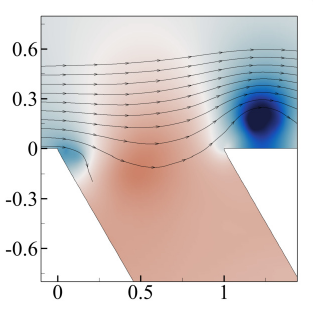}};
	\begin{scope}[x={(image.south east)},y={(image.north west)}]
		\fill [white] (0.0,0.9) rectangle (0.09,1.0);
	\end{scope}
\end{tikzpicture}
\begin{tikzpicture}
	\node[anchor=south west,inner sep=0] (image) at (0,0) {\includegraphics[width=0.29\textwidth]{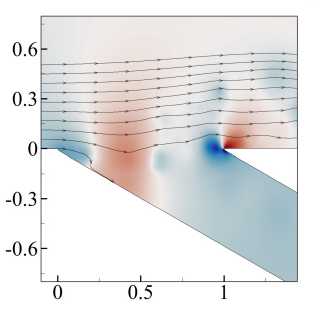}};
	\begin{scope}[x={(image.south east)},y={(image.north west)}]
		\fill [white] (0.0,0.9) rectangle (0.09,1.0);
	\end{scope}
\end{tikzpicture}
    \hspace{5cm}
\begin{tikzpicture}
	\node[anchor=south west,inner sep=0] (image) at (0,0) {\includegraphics[width=0.29\textwidth]{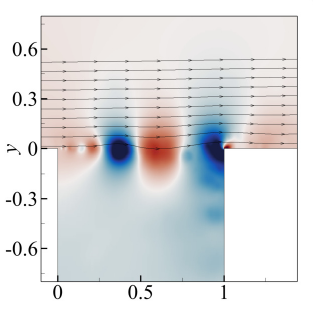}};
	\begin{scope}[x={(image.south east)},y={(image.north west)}]
		\fill [white] (0.0,0.9) rectangle (0.09,1.0);
		\node[anchor=north west, inner sep=1pt] at (0,1) {(\emph{c})};
	\end{scope}
\end{tikzpicture}
\begin{tikzpicture}
	\node[anchor=south west,inner sep=0] (image) at (0,0) {\includegraphics[width=0.29\textwidth]{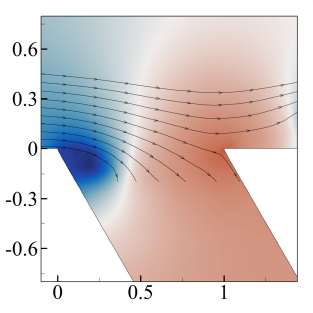}};
	\begin{scope}[x={(image.south east)},y={(image.north west)}]
		\fill [white] (0.0,0.9) rectangle (0.09,1.0);
	\end{scope}
\end{tikzpicture}
\begin{tikzpicture}
	\node[anchor=south west,inner sep=0] (image) at (0,0) {\includegraphics[width=0.29\textwidth]{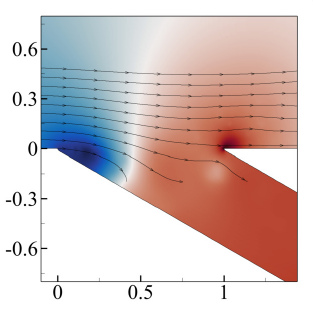}};
	\begin{scope}[x={(image.south east)},y={(image.north west)}]
		\fill [white] (0.0,0.9) rectangle (0.09,1.0);
	\end{scope}
\end{tikzpicture}
    \hspace{5cm}
\begin{tikzpicture}
	\node[anchor=south west,inner sep=0] (image) at (0,0) {\includegraphics[width=0.29\textwidth]{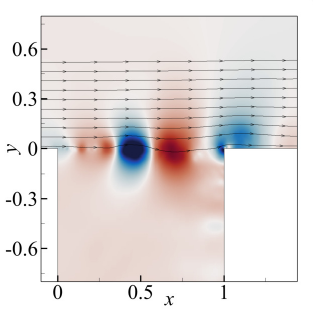}};
	\begin{scope}[x={(image.south east)},y={(image.north west)}]
		\fill [white] (0.0,0.9) rectangle (0.09,1.0);
		\node[anchor=north west, inner sep=1pt] at (0,1) {(\emph{d})};
	\end{scope}
\end{tikzpicture}
\begin{tikzpicture}
	\node[anchor=south west,inner sep=0] (image) at (0,0) {\includegraphics[width=0.29\textwidth]{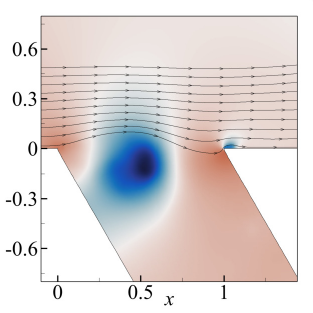}};
	\begin{scope}[x={(image.south east)},y={(image.north west)}]
		\fill [white] (0.0,0.9) rectangle (0.09,1.0);
	\end{scope}
\end{tikzpicture}
\begin{tikzpicture}
	\node[anchor=south west,inner sep=0] (image) at (0,0) {\includegraphics[width=0.29\textwidth]{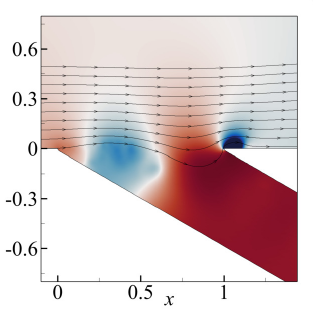}};
	\begin{scope}[x={(image.south east)},y={(image.north west)}]
		\fill [white] (0.0,0.9) rectangle (0.09,1.0);
	\end{scope}
\end{tikzpicture}
    \caption{Snapshots of the spanwise-averaged instantaneous hydrodynamic pressure fluctuation $p'_{H}$ with superimposed streamlines to signify the shear-layer undulation across the cavity opening with a time interval of $T$/4 between two successive plots from (\emph{a}) to (\emph{d}), where $T$ is the period of the oscillation cycle of $\chi$. The first, second and third columns correspond to $\alpha=90^{\circ}$, $60^{\circ}$ and $30^{\circ}$, respectively.}
	\label{fig:Fig12}
\end{figure}
\begin{figure}[!h]
	\centering
\begin{tikzpicture}
	\node[anchor=south west,inner sep=0] (image) at (0,0) {\includegraphics[width=0.29\textwidth]{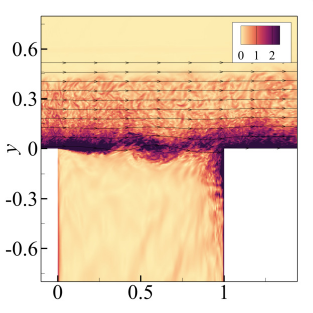}};
	\begin{scope}[x={(image.south east)},y={(image.north west)}]
		\fill [white] (0.0,0.9) rectangle (0.09,1.0);
		\node[anchor=north west, inner sep=1pt] at (0,1) {(\emph{a})};
	\end{scope}
\end{tikzpicture}
\begin{tikzpicture}
	\node[anchor=south west,inner sep=0] (image) at (0,0) {\includegraphics[width=0.29\textwidth]{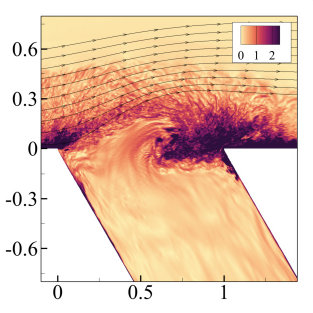}};
	\begin{scope}[x={(image.south east)},y={(image.north west)}]
		\fill [white] (0.0,0.9) rectangle (0.09,1.0);
	\end{scope}
\end{tikzpicture}
\begin{tikzpicture}
	\node[anchor=south west,inner sep=0] (image) at (0,0) {\includegraphics[width=0.29\textwidth]{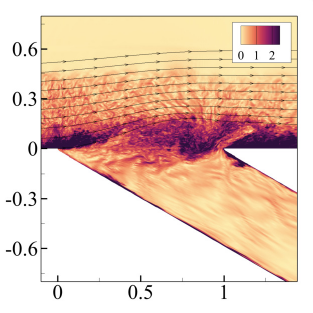}};
	\begin{scope}[x={(image.south east)},y={(image.north west)}]
		\fill [white] (0.0,0.9) rectangle (0.09,1.0);
	\end{scope}
\end{tikzpicture}
    \hspace{5cm}
\begin{tikzpicture}
	\node[anchor=south west,inner sep=0] (image) at (0,0) {\includegraphics[width=0.29\textwidth]{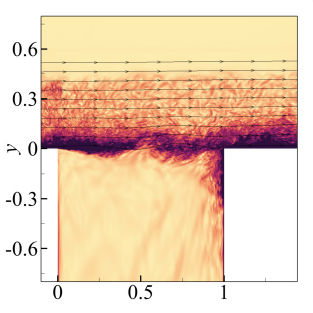}};
	\begin{scope}[x={(image.south east)},y={(image.north west)}]
		\fill [white] (0.0,0.9) rectangle (0.09,1.0);
		\node[anchor=north west, inner sep=1pt] at (0,1) {(\emph{b})};
	\end{scope}
\end{tikzpicture}
\begin{tikzpicture}
	\node[anchor=south west,inner sep=0] (image) at (0,0) {\includegraphics[width=0.29\textwidth]{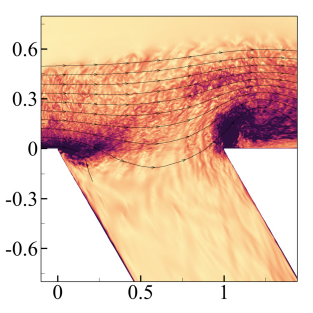}};
	\begin{scope}[x={(image.south east)},y={(image.north west)}]
		\fill [white] (0.0,0.9) rectangle (0.09,1.0);
	\end{scope}
\end{tikzpicture}
\begin{tikzpicture}
	\node[anchor=south west,inner sep=0] (image) at (0,0) {\includegraphics[width=0.29\textwidth]{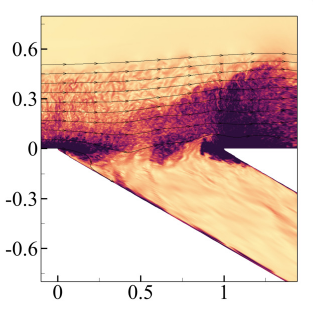}};
	\begin{scope}[x={(image.south east)},y={(image.north west)}]
		\fill [white] (0.0,0.9) rectangle (0.09,1.0);
	\end{scope}
\end{tikzpicture}
    \hspace{5cm}
\begin{tikzpicture}
	\node[anchor=south west,inner sep=0] (image) at (0,0) {\includegraphics[width=0.29\textwidth]{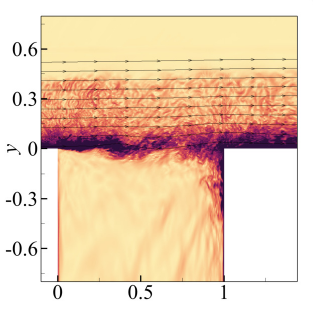}};
	\begin{scope}[x={(image.south east)},y={(image.north west)}]
		\fill [white] (0.0,0.9) rectangle (0.09,1.0);
		\node[anchor=north west, inner sep=1pt] at (0,1) {(\emph{c})};
	\end{scope}
\end{tikzpicture}
\begin{tikzpicture}
	\node[anchor=south west,inner sep=0] (image) at (0,0) {\includegraphics[width=0.29\textwidth]{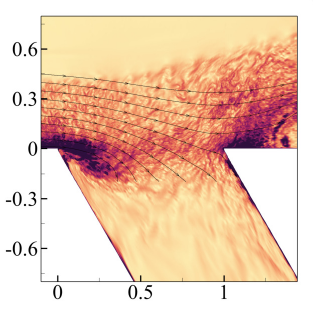}};
	\begin{scope}[x={(image.south east)},y={(image.north west)}]
		\fill [white] (0.0,0.9) rectangle (0.09,1.0);
	\end{scope}
\end{tikzpicture}
\begin{tikzpicture}
	\node[anchor=south west,inner sep=0] (image) at (0,0) {\includegraphics[width=0.29\textwidth]{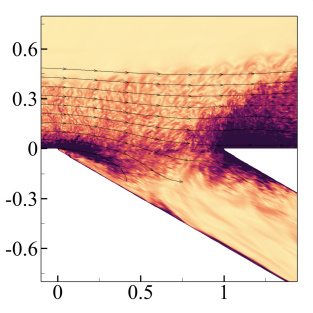}};
	\begin{scope}[x={(image.south east)},y={(image.north west)}]
		\fill [white] (0.0,0.9) rectangle (0.09,1.0);
	\end{scope}
\end{tikzpicture}
    \hspace{5cm}
\begin{tikzpicture}
	\node[anchor=south west,inner sep=0] (image) at (0,0) {\includegraphics[width=0.29\textwidth]{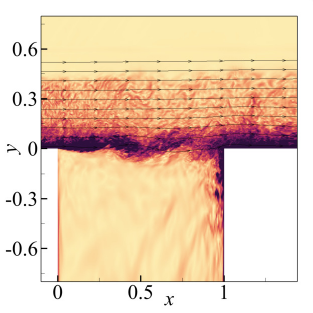}};
	\begin{scope}[x={(image.south east)},y={(image.north west)}]
		\fill [white] (0.0,0.9) rectangle (0.09,1.0);
		\node[anchor=north west, inner sep=1pt] at (0,1) {(\emph{d})};
	\end{scope}
\end{tikzpicture}
\begin{tikzpicture}
	\node[anchor=south west,inner sep=0] (image) at (0,0) {\includegraphics[width=0.29\textwidth]{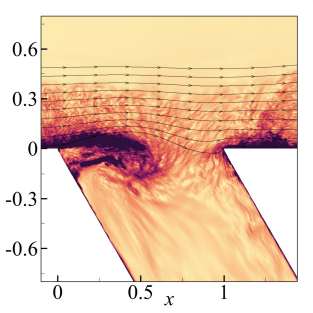}};
	\begin{scope}[x={(image.south east)},y={(image.north west)}]
		\fill [white] (0.0,0.9) rectangle (0.09,1.0);
	\end{scope}
\end{tikzpicture}
\begin{tikzpicture}
	\node[anchor=south west,inner sep=0] (image) at (0,0) {\includegraphics[width=0.29\textwidth]{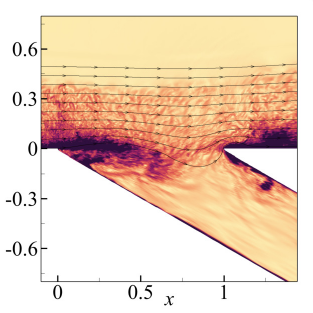}};
	\begin{scope}[x={(image.south east)},y={(image.north west)}]
		\fill [white] (0.0,0.9) rectangle (0.09,1.0);
	\end{scope}
\end{tikzpicture}
    \caption{Snapshots of the spanwise-averaged instantaneous DLE with superimposed streamlines to signify the shear-layer undulation across the cavity opening with a time interval of $T$/4 between two successive plots from (\emph{a}) to (\emph{d}), where $T$ is the period of the oscillation cycle of $\chi$. The first, second and third columns correspond to $\alpha=90^{\circ}$, $60^{\circ}$ and $30^{\circ}$, respectively. For the corresponding Q-criterion fields, refer to figure \ref{fig:Fig11}.}
	\label{fig:Fig13}
\end{figure}

Figure \ref{fig:Fig11} shows an oscillation cycle of $\chi$ similar to that of figure \ref{fig:Fig6}, with particular attention now given to the vortex dynamics near the cavity opening region. The $Q$-criterion, calculated from Eq. (\ref{eq:div2p}), is plotted and superimposed with streamlines to indicate the shear-layer oscillations near the cavity opening. For brevity, we examine the vortex dynamics within the $\alpha=60^{\circ}$ inclined cavity. Figure \ref{fig:Fig11}(\emph{a}) shows the instant when a large-scale vortex, characterized by $Q<0$, is located slightly above the cavity opening line at the downstream corner. The vorticity-dominated region near the downstream walls is associated with low hydrodynamic pressure ($p'_{H}<0$) and is visualized in figure \ref{fig:Fig12}. These hydrodynamic wall-pressure fluctuations act as a dipole noise source from the surface integral of Curle's equation \cite{curle1955influence}, which were previously identified as the dominant noise source in both shallow and deep cavities \cite{larsson2004aeroacoustic,lam2013aeroacoustics}. Therefore, the presence of low hydrodynamic wall-pressure fluctuations ($p'_H<0$) correspond to generating rarefaction acoustic waves ($p'_A<0$), which at this point destructively interfere with compressive acoustic waves ($p'_A>0$) reflected from the cavity base during the preceding cycle. This destructive interference leads to the acoustic pressure equilibrium within the cavity (i.e., $\chi=0$), as previously discussed in figure \ref{fig:Fig6}(\emph{a}).

Continuing to examine the $\alpha=60^\circ$ case, as the large-scale vortex sweeps past the downstream corner, it induces stronger interaction with the vorticity field and the downstream walls. At this stage, the additional rarefaction waves undergo constructive interference with the acoustic waves reflected from the base of the cavity, until the large-scale vortex is completely ejected from the cavity, as shown in figure \ref{fig:Fig11}(\emph{b}). The ejection of the vortex triggers an immediate downward flapping of the shear-layer near the upstream corner, accompanied by the emergence of small-scale vortices within the separated shear-layer. These small vortices gradually grow and merge, eventually forming a single large-scale vortex near the upstream corner, as illustrated in figure \ref{fig:Fig11}(\emph{c}). The further downstream convection and development of this large-scale vortex before completing the feedback oscillation cycle is shown in figure \ref{fig:Fig11}(\emph{d}). In contrast to the orthogonal cavity case, the recurring interaction between the escaping vortical structure and the downstream corner in inclined cavities mirrors the well-documented phenomenon of ``vortex above corner'' interaction, as reported by Tang and Rockwell \cite{Tang1983}.

Additionally, we employ the Direct Lyapunov Exponents (DLE) method introduced by Haller \cite{HallerYuan2000, Green2006} to identify the Lagrangian Coherent Structures (LCS) present in the turbulent cavity flow fields. The DLE field is defined as
\begin{equation}\label{eq:DLE}
    \text{DLE}_{\Delta T}(x_0, t_0) = \frac{1}{2\Delta T} \log(\sigma_T(x_0, t_0)),
\end{equation}
where $\sigma_T$ represents the square of the largest singular value of the Cauchy-Green deformation tensor such that
\begin{equation}\label{eq:DLEsigma}
    \sigma_T(x_0, t_0) = \lambda_{\max} \left( \left[ \frac{\partial x(t_0 + \Delta T, x_0, t_0)}{\partial x_0} \right]^T \left[ \frac{\partial x(t_0 + \Delta T, x_0, t_0)}{\partial x_0} \right] \right),
\end{equation}
and $x(t, x_0, t_0)$ denotes the position of a particle at time \( t \), initiating at position $x_0$ at time $t_0$. The DLE field is calculated by integrating trajectories in backward time ($\Delta T < 0$) and the ridges in the DLE field capture attracting Lagrangian coherent structures (attracting LCS) in the flow field \cite{HallerYuan2000}. The integration time, $\Delta T$, is adjusted to achieve the desired level of detail in the calculation without compromising the location of the attracting LCS boundary. The DLE field is visualised in figure \ref{fig:Fig13}.

Based on figures \ref{fig:Fig11} -- \ref{fig:Fig13} (also video clips provided as a supplementary material to this paper), we highlight that there is a distinctive flow mechanism in the inclined cavities which is significantly different from that of the orthogonal cavity. Firstly, the inclined cavities produce a single large-scale vortical structure travelling across the cavity opening which manifests a 1st hydrodynamic mode. In contrast, the orthogonal cavity exhibits two smaller vortices moving simultaneously along the cavity, corresponding to a 2nd hydrodynamic mode. In the inclined cavities, there is a pronounced level of Kelvin-Helmholtz (K-H) instability and a flapping motion of the shear layer. When the shear layer flaps downward, an intense roll-up vortex forms at the upstream corner. This vortex continues to grow in size by merging with nearby smaller eddies as the shear layer flaps upward. Consequently, the vortex spends a substantial amount of time growing rather than rapidly traveling downstream (as will be shown later), resulting in a significantly longer cycle period. Due to its large size, only one vortex occupies the cavity, as opposed to two in the orthogonal case. We suggest that these are the main reasons why the inclined cavities exhibit the 1st hydrodynamic mode which selects a lower-frequency resonance with the 1st acoustic mode. Additionally, the relatively higher acoustic pulsation energy contained within the inclined cavities (as deduced from the eigenmode analysis in the previous section) may have reinforced the excitation of K-H instability and shear-layer flapping.

\begin{figure}[!h]
	\centering
\begin{tikzpicture}
	\node[anchor=south west,inner sep=0] (image) at (0,0) {\includegraphics[width=0.75\textwidth]{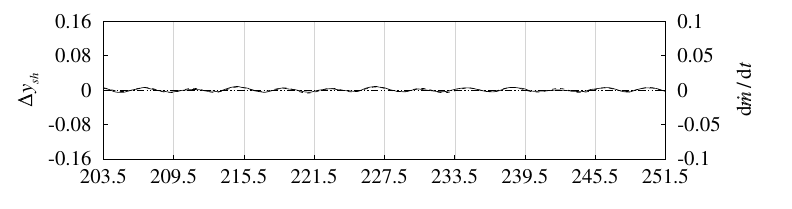}};
	\begin{scope}[x={(image.south east)},y={(image.north west)}]
		\fill [white] (0.0,0.95) rectangle (0.09,1.0);
		\node[anchor=north west, inner sep=5pt] at (0,1) {(\emph{a})};
	\end{scope}
\end{tikzpicture}
\begin{tikzpicture}
	\node[anchor=south west,inner sep=0] (image) at (0,0) {\includegraphics[width=0.75\textwidth]{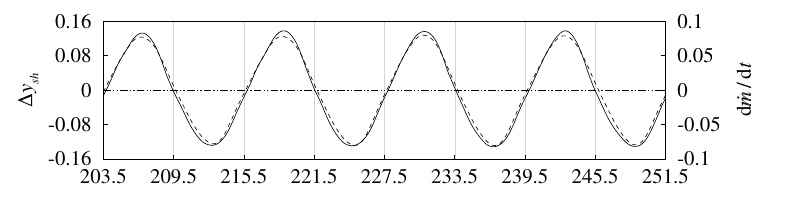}};
	\begin{scope}[x={(image.south east)},y={(image.north west)}]
		\fill [white] (0.0,0.95) rectangle (0.09,1.0);
		\node[anchor=north west, inner sep=5pt] at (0,1) {(\emph{b})};
	\end{scope}
\end{tikzpicture}
\begin{tikzpicture}
	\node[anchor=south west,inner sep=0] (image) at (0,0) {\includegraphics[width=0.75\textwidth]{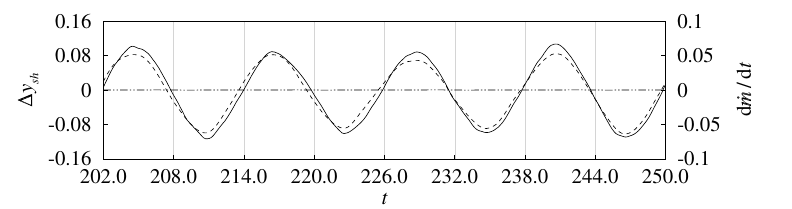}};
	\begin{scope}[x={(image.south east)},y={(image.north west)}]
		\fill [white] (0.0,0.95) rectangle (0.09,1.0);
		\node[anchor=north west, inner sep=5pt] at (0,1) {(\emph{c})};
	\end{scope}
\end{tikzpicture}
\caption{Time histories of (\dashed) $d\dot{m}_A/dt$, (\chain) $d\dot{m}_H/dt$ and (\full) $\Delta y_{sh}$ for (\emph{a}) $\alpha=90^{\circ}$, (\emph{b}) $60^{\circ}$ and (\emph{c}) $30^{\circ}$, respectively.}
\label{fig:Fig14}
\end{figure}

Accordingly, the shear-layer flapping may be quantified by the time rate of change of mass flow rate through the cavity opening. First, based on equation \ref{eq:doak0} used in MPT, the total mass flow rate is decomposed into hydrodynamic ($\dot{m}_H$) and acoustic ($\dot{m}_A$) components, and their time rate of change are expressed as
\begin{subequations}
	\begin{align}
        \frac{d\dot{m_{H}}}{dt}(t)&=\frac{1}{dt}\int_0^1{(\rho v)_H|_{y=0}}dx=\frac{1}{dt}\int_{0}^{1}{B_{y}|_{y=0}dx},\\ 
		\frac{d\dot{m_{A}}}{dt}(t)&=\frac{1}{dt}\int_0^1{(\rho v)_A|_{y=0}dx}=-\frac{1}{dt}\int_{0}^{1}{\left.\frac{\partial\psi_A}{\partial y}\right|_{y=0}dx}.
	\end{align}
    \label{eq:dmadtanddmhdt}
\end{subequations}
In addition, the shear-layer flapping may also be characterised by the displacement of a streamline (emerging from the upstream corner), $\Delta y_{sh}$, measured at the center of the cavity opening, $(x,y)=(0.5,0)$. Figure \ref{fig:Fig14} shows the time history of $d\dot{m}_H/dt$, $d\dot{m}_A/dt$ and $\Delta y_{sh}$ for each cavity. It is apparent that energetic shear-layer flapping is present in the inclined cavities and is almost perfectly synchronized with $d\dot{m}_A/dt$ that is induced by the depthwise acoustic resonance. Conversely, the contribution from the hydrodynamic component ($\dot{m}_H$) remains effectively zero, in accordance with the requirement that the net mass flow rate of solenoidal (incompressible) flow in a confined geometry should remain zero. Nonetheless, the hydrodynamic component of the vertical momentum $(\rho v)_H$ provides essential insight into the vortex dynamics developing across the cavity opening, which will be discussed next.

\begin{figure}[!h]
	\centering
\begin{tikzpicture}
	\node[anchor=south west,inner sep=0] (image) at (0,0) {\includegraphics[width=0.30\textwidth]{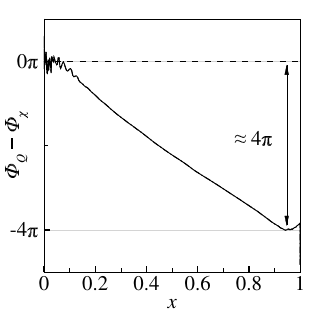}};
	\begin{scope}[x={(image.south east)},y={(image.north west)}]
		\fill [white] (0.0,0.9) rectangle (0.09,1.0);
		\node[anchor=north west, inner sep=5pt] at (0,1) {(\emph{a})};
	\end{scope}
\end{tikzpicture}
\begin{tikzpicture}
	\node[anchor=south west,inner sep=0] (image) at (0,0) {\includegraphics[width=0.30\textwidth]{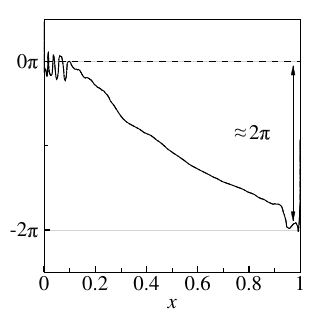}};
	\begin{scope}[x={(image.south east)},y={(image.north west)}]
		\fill [white] (0.0,0.9) rectangle (0.09,1.0);
		\node[anchor=north west, inner sep=5pt] at (0,1) {(\emph{b})};
	\end{scope}
\end{tikzpicture}
\begin{tikzpicture}
	\node[anchor=south west,inner sep=0] (image) at (0,0) {\includegraphics[width=0.30\textwidth]{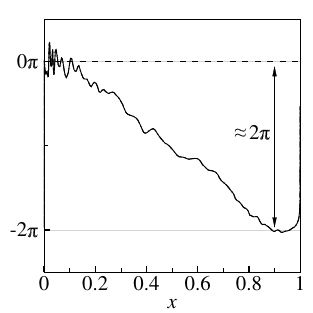}};
	\begin{scope}[x={(image.south east)},y={(image.north west)}]
		\fill [white] (0.0,0.9) rectangle (0.09,1.0);
		\node[anchor=north west, inner sep=5pt] at (0,1) {(\emph{c})};
	\end{scope}
\end{tikzpicture}
	\caption{
		The streamwise phase variation of the Fourier-transformed $Q$-criterion, $\Phi_{Q}(\xx,f)$, measured across the cavity opening (i.e., at $y=0$) at their respective tonal frequencies for (\emph{a}) $\alpha=90^{\circ}$, (\emph{b}) $60^{\circ}$ and (\emph{c}) $30^{\circ}$, respectively. Note that the spatial variation of $\Phi_{Q}(\xx,f)$ is calculated based on the phase reference of $\Phi_{\chi}(\xx,f)$. The plots reveal regions of frequency modulation where the $Q$-Criterion fluctuation near the upstream and downstream corners remains highly synchronized with the averaged acoustic wall-pressure fluctuation at the cavity base ($\cos[\Phi_{Q}(\xx,f)-\Phi_{\chi}(\xx,f)]=1$). This observation suggests that the phase of hydrodynamic cycle (e.g., initial vortex formation and subsequent impingement) is highly synchronized with the acoustic cycle (i.e., depthwise acoustic resonance).}
	\label{fig:Fig15}
\end{figure}

Figure \ref{fig:Fig15} shows streamwise phase variations of the $Q$-criterion, $\Phi_{Q}(\xx,f)$, across the cavity opening for both orthogonal and inclined cavities. The phase variations confirm the 2nd ($n=2$) and 1st ($n=1$) hydrodynamic modes present in the orthogonal and inclined cavities, respectively, which satisfy the phase condition $\Delta \Phi_{Q}(\xx,f) = 2\pi n$, as detailed by Rockwell and Naudascher \cite{Rockwell1979}. The plots also reveal that the hydrodynamic fluctuations near the upstream and downstream corners remain highly coherent with the wall pressure fluctuation at the base of the cavity ($\cos[\Phi_{Q}(\xx,f)-\Phi_{\chi}(\xx,f)]=1$). This indicates that the phase of the hydrodynamic cycle (from initial vortex formation through final scattering) is synchronised with the acoustic cycle (i.e. depthwise acoustic resonance), emphasising the lock-in condition.

\begin{figure}[!h]
	\centering
	\begin{tikzpicture}
		\node[anchor=south west,inner sep=0] (image) at (0,0) {\includegraphics[width=0.285\textwidth]{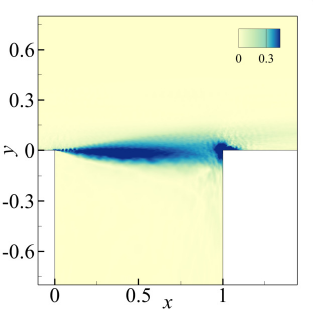}};
		\begin{scope}[x={(image.south east)},y={(image.north west)}]
			\fill [white] (0.0,0.9) rectangle (0.09,1.0);
			\node[anchor=north west, inner sep=0pt] at (0,1) {(\emph{a})};
		\end{scope}
	\end{tikzpicture}
	\begin{tikzpicture}
		\node[anchor=south west,inner sep=0] (image) at (0,0) {\includegraphics[width=0.285\textwidth]{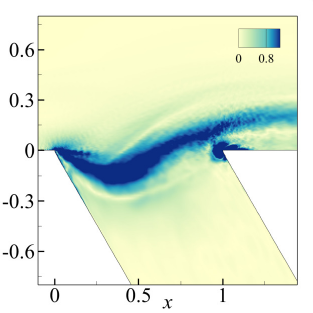}};
		\begin{scope}[x={(image.south east)},y={(image.north west)}]
			\fill [white] (0.0,0.9) rectangle (0.09,1.0);
			\node[anchor=north west, inner sep=0pt] at (0,1) {(\emph{b})};
		\end{scope}
	\end{tikzpicture}
	\begin{tikzpicture}
		\node[anchor=south west,inner sep=0] (image) at (0,0) {\includegraphics[width=0.285\textwidth]{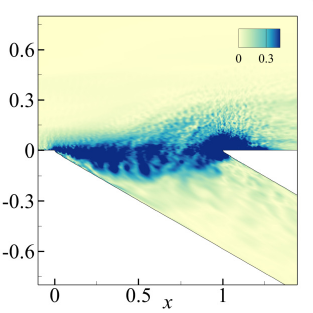}};
		\begin{scope}[x={(image.south east)},y={(image.north west)}]
			\fill [white] (0.0,0.9) rectangle (0.09,1.0);
			\node[anchor=north west, inner sep=0pt] at (0,1) {(\emph{c})};
		\end{scope}
	\end{tikzpicture}
	\caption{The contour plots show the spatial distribution of the magnitude of Fourier-transformed $Q$-criterion, $|Q(\xx,f)|$ at their respective tonal frequencies for (\emph{a}) $\alpha=90^{\circ}$, (\emph{b}) $60^{\circ}$ and (\emph{c}) $30^{\circ}$, respectively.}
	\label{fig:Fig16}
\end{figure}

In order to examine the hydrodynamic field in more detail, a Fourier transform of $Q$-criterion is calculated and its magnitude at the resonance frequency is plotted in figure \ref{fig:Fig16}. This plot reveals the zones of major vortical activity contributing to the resonance cycle in each of the three cavity geometries. As expected, the inclined cavities display broader zones of activity, reflecting the larger vortex structures observed earlier. In the $\alpha=60^\circ$ case, the vortex is more intense and traces a distinctly undulating path, yet it impinges less directly on the downstream corner. By contrast, the less intense vortices at $\alpha=30^\circ$ strike the corner more directly. These contrasting behaviors likely balance one another, leading to comparable overall sound levels in both inclined cavities. However, it is important to emphasize that these insights pertain solely to the single Fourier component at $St = 0.276$ and do not capture the full spectral dynamics of the flow.

\begin{figure}[!h]
	\centering
\begin{tikzpicture}
	\node[anchor=south west,inner sep=0] (image) at (0,0) {\includegraphics[width=0.8\textwidth]{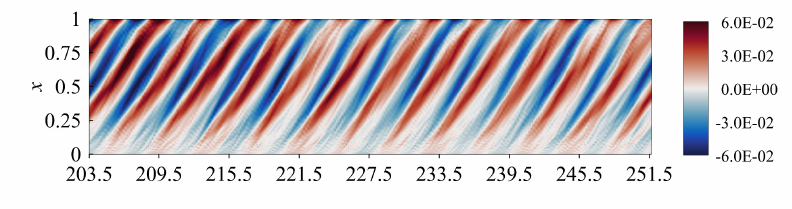}};
	\begin{scope}[x={(image.south east)},y={(image.north west)}]
		\fill [white] (0.0,0.85) rectangle (0.09,1.0);
		\node[anchor=north west, inner sep=3pt] at (0,1) {(\emph{a})};
	\end{scope}
\end{tikzpicture}
\begin{tikzpicture}
	\node[anchor=south west,inner sep=0] (image) at (0,0) {\includegraphics[width=0.8\textwidth]{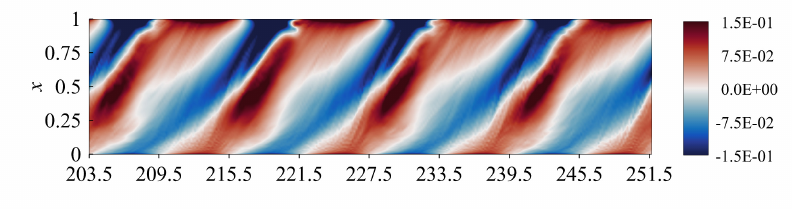}};
	\begin{scope}[x={(image.south east)},y={(image.north west)}]
		\fill [white] (0.0,0.85) rectangle (0.09,1.0);
		\node[anchor=north west, inner sep=3pt] at (0,1) {(\emph{b})};
	\end{scope}
\end{tikzpicture}
\begin{tikzpicture}
	\node[anchor=south west,inner sep=0] (image) at (0,0) {\includegraphics[width=0.8\textwidth]{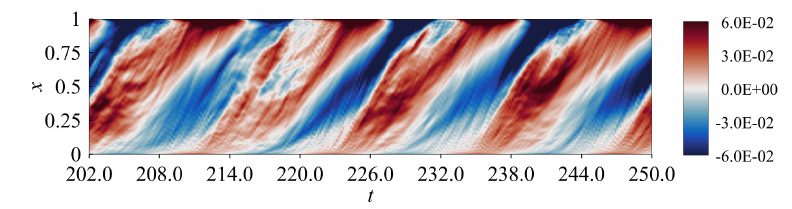}};
	\begin{scope}[x={(image.south east)},y={(image.north west)}]
		\fill [white] (0.0,0.85) rectangle (0.09,1.0);
		\node[anchor=north west, inner sep=3pt] at (0,1) {(\emph{c})};
	\end{scope}
\end{tikzpicture}
    \caption{Spatio-temporal contour plots of the hydrodynamic component of the vertical momentum $(\rho v)_H$ across the cavity opening ($y=0$) for (\emph{a}) $\alpha=90^\circ$, (\emph{b}) $60^\circ$ and (\emph{c}) $30^\circ$, respectively.}
	\label{fig:Fig17}
\end{figure}

\begin{figure}[!h]
	\centering
    {\includegraphics[width=0.7\textwidth]{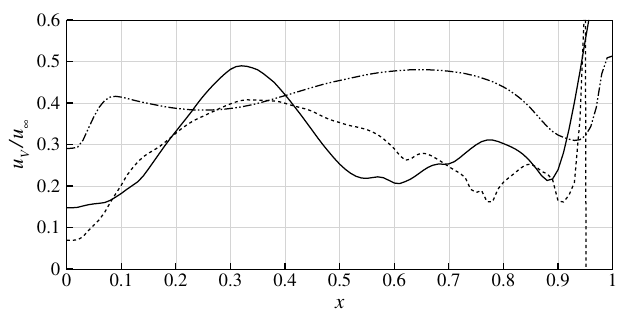}}
	\caption{
        Representative vortex travel speed $u_V/u_\infty$ estimated for (\chain) $\alpha=90^{\circ}$, (\full) $60^{\circ}$ and (\dashed) $30^{\circ}$, respectively. This estimation is based on iso-contour lines of $(\rho v)_H=0$ in figure \ref{fig:Fig17} from which $u_V=dx/dt$ is calculated.}
        \label{fig:Fig18}
\end{figure}

\begin{figure}[!h]
	\centering
    {\includegraphics[width=0.35\textwidth]{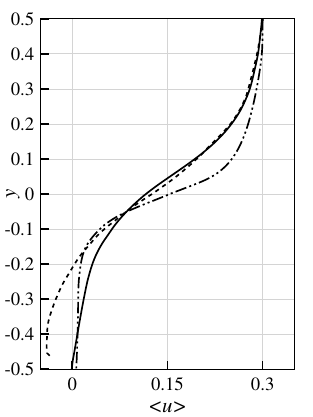}}
	\caption{Time-averaged profiles of streamwise velocity across the shear layer measured at $x = 0.8$ for (\chain) $\alpha=90^{\circ}$, (\full) $60^{\circ}$ and (\dashed) $30^{\circ}$, displaying thicker shear layers produced in the inclined cavities compared to the orthogonal case.} \label{fig:Fig19}
\end{figure}

We discussed earlier about the relatively slower travel speed of the large vortex in the inclined cavities in relation to the growth process of the vortex that is not particularly featured in the orthogonal cavity. In order to estimate the speed of the vortex travel, we look at the hydrodynamic component of the vertical momentum $(\rho v)_H$ fluctuating in space and time (across the cavity opening), as shown in figure \ref{fig:Fig17}. We suggest that the spatio-temporal contour plot of $(\rho v)_H$ (which contains purely hydrodynamic fluctuations) displays the footprint of the large-scale vortex structure moving across the cavity opening. If we take an iso-contour line for $(\rho v)_H=0$ (one of the white curves in figure \ref{fig:Fig17}) and calculate the slope of the curve, i.e. $dx/dt$ as a function of $x$, it shows a representative speed of the vortex travel ($u_V$) at each point across the cavity opening ($x\in[0,1]$ and $y=0$). This is repeated for 10 consecutive curves of $(\rho v)_H=0$ and an averaged is obtained. Figure \ref{fig:Fig18} shows that $u_V$ is significantly lower in the inclined cavities than that of the orthogonal cavity across most of the cavity opening. Furthermore, it is apparent that the slower vortex travel of both inclined cavities occurs in two separate regions: (1) near the upstream corner $x\in[0,0.2]$ and (2) in the mid-to-downstream region $x\in[0.4,0.9]$. We suggest that the former is linked with the initial roll-up of the vortex during the downward flapping of the shear layer, and the latter with the further growth of the vortex by entraining/merging adjacent eddies during the upward flapping of the shear layer. Here, an additional observation is that the inclined cavities, owing to their unique vortex dynamics, produce thicker shear layers than the orthogonal cavity case, as shown in figure \ref{fig:Fig19}. In this regard, Mathias and Medeiros \cite{Mathias2021} recently investigated the stability of cavity mixing/shear layers. They found that the most unstable mode frequency of a shear layer decreases as its thickness increases. This earlier study (although conducted for shallow and orthogonal cavities) supports the current explanation of the slower vortex travel sustained by the lower-frequency instability of the thicker shear layers in the inclined cavities.

\section{Aeroacoustic mode amplifications and selections}\label{sec:inoutputanalysis}
Section \ref{sec:hydrodynamicfield} highlighted that the vortex dynamics plays a key role in determining the acoustic resonance mode and amplitude in deep and inclined cavities. Classical aeroacoustic theories that involve vortex dynamics \cite{Howe1998, Howe2003} suggest that instantaneous acoustic source power, $\Pi$, can be approximated using the following expression
\begin{equation}\label{eq:HEC}
	\Pi \approx -\rho_{\infty} \iint_{\boldsymbol{x}} (\boldsymbol{\omega} \times \boldsymbol{u}) \cdot \boldsymbol{u}_{a} \; d\boldsymbol{x},
\end{equation}
where $\boldsymbol{\omega}$ is the vorticity vector and $\boldsymbol{u}_{a}$ refers to the acoustic particle velocity vector. The triple-dot product ($(\boldsymbol{\omega} \times \boldsymbol{u}) \cdot \boldsymbol{u}_{a}$) provides quantitative insight into the local energy transfer between the hydrodynamic and acoustic fields. In particular, the integrand captures the transfer of acoustic energy to hydrodynamic energy (e.g., $ (\boldsymbol{\omega} \times \boldsymbol{u}) \cdot \boldsymbol{u}_{a} > 0$) and vice versa. For sustained oscillations to occur, it is important for the integral in Eq. (\ref{eq:HEC}) to remain positive over an acoustic cycle, ensuring that a favorable phase relationship between the Lamb vector ($\boldsymbol{\omega} \times \boldsymbol{u}$) and the acoustic field, $\boldsymbol{u}_{a}$. The following discussion examines this energy exchange process by analyzing the temporal dynamics of the Lamb vector within resonant acoustic fields in inclined cavities.

\begin{figure}[!h]
	\centering
\begin{tikzpicture}
	\node[anchor=south west,inner sep=0] (image) at (0,0) {\includegraphics[width=0.3\textwidth]{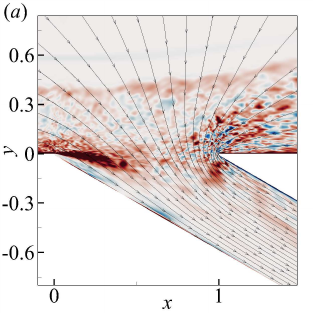}};
	\begin{scope}[x={(image.south east)},y={(image.north west)}]
		\fill [white] (0.0,0.9) rectangle (0.09,1.0);
		\fill [white] (0.45,0.0) rectangle (0.55,0.075);
		\node[anchor=north west, inner sep=0pt] at (0,1) {(\emph{a})};
	\end{scope}
\end{tikzpicture}
\begin{tikzpicture}
	\node[anchor=south west,inner sep=0] (image) at (0,0) {\includegraphics[width=0.3\textwidth]{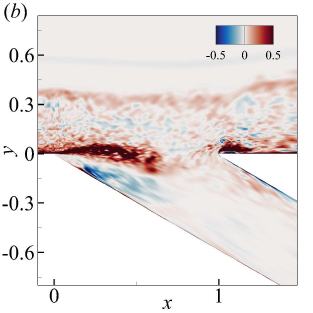}};
	\begin{scope}[x={(image.south east)},y={(image.north west)}]
		\fill [white] (0.0,0.9) rectangle (0.09,1.0);
		\fill [white] (0.0,0.45) rectangle (0.075,0.55);
		\fill [white] (0.45,0.0) rectangle (0.55,0.075);
		\node[anchor=north west, inner sep=0pt] at (0,1) {(\emph{b})};
	\end{scope}
\end{tikzpicture}
	\hspace{5cm}
\begin{tikzpicture}
	\node[anchor=south west,inner sep=0] (image) at (0,0) {\includegraphics[width=0.3\textwidth]{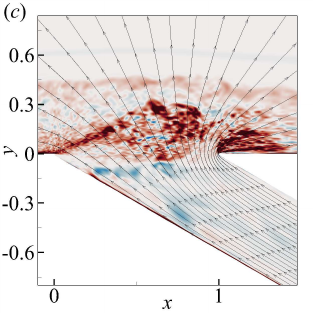}};
	\begin{scope}[x={(image.south east)},y={(image.north west)}]
		\fill [white] (0.0,0.9) rectangle (0.09,1.0);
		\node[anchor=north west, inner sep=0pt] at (0,1) {(\emph{c})};
	\end{scope}
\end{tikzpicture}
\begin{tikzpicture}
	\node[anchor=south west,inner sep=0] (image) at (0,0) {\includegraphics[width=0.3\textwidth]{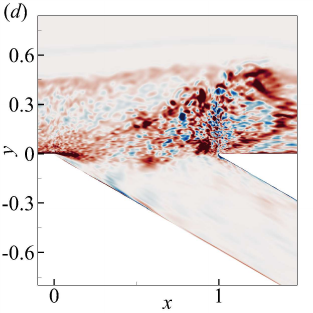}};
	\begin{scope}[x={(image.south east)},y={(image.north west)}]
		\fill [white] (0.0,0.9) rectangle (0.09,1.0);
		\fill [white] (0.0,0.45) rectangle (0.075,0.55);
		\node[anchor=north west, inner sep=0pt] at (0,1) {(\emph{d})};
	\end{scope}
\end{tikzpicture}
    \caption{The time evolution of the Lamb vector ($\boldsymbol{\omega} \times \boldsymbol{u}$) in the vertical direction for the $\alpha=30^{\circ}$ inclined cavity is examined for a single acoustic cycle. The contour plots capture two key time instants: those (\emph{a}) when the instantaneous acoustic source power $\Pi$ reaches its minimum and (\emph{c}) when it reaches its maximum. The plots (\emph{b}) and (\emph{d}) indicate the time junctures when the instantaneous acoustic source power becomes zero, such as when $\Pi=0$. Here, the superimposed streamline represents the instantaneous acoustic particle velocity field.}
	\label{fig:Fig20}
\end{figure}
\begin{figure}[!h]
	\centering
	\begin{tikzpicture}
		\node[anchor=south west,inner sep=0] (image) at (0,0) {\includegraphics[width=0.3\textwidth]{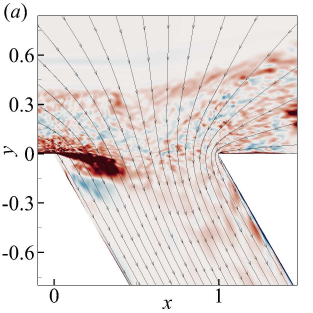}};
		\begin{scope}[x={(image.south east)},y={(image.north west)}]
			\fill [white] (0.0,0.9) rectangle (0.09,1.0);
    		\fill [white] (0.45,0.0) rectangle (0.55,0.075);
			\node[anchor=north west, inner sep=0pt] at (0,1) {(\emph{a})};
		\end{scope}
	\end{tikzpicture}
	\begin{tikzpicture}
		\node[anchor=south west,inner sep=0] (image) at (0,0) {\includegraphics[width=0.3\textwidth]{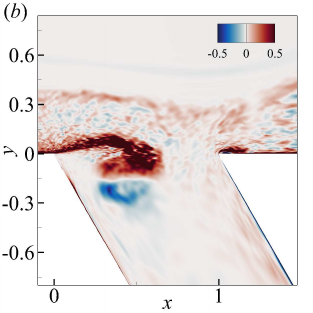}};
		\begin{scope}[x={(image.south east)},y={(image.north west)}]
			\fill [white] (0.0,0.9) rectangle (0.09,1.0);
    		\fill [white] (0.0,0.45) rectangle (0.075,0.55);
    		\fill [white] (0.45,0.0) rectangle (0.55,0.075);
			\node[anchor=north west, inner sep=0pt] at (0,1) {(\emph{b})};
		\end{scope}
	\end{tikzpicture}
	\hspace{5cm}
	\begin{tikzpicture}
		\node[anchor=south west,inner sep=0] (image) at (0,0) {\includegraphics[width=0.3\textwidth]{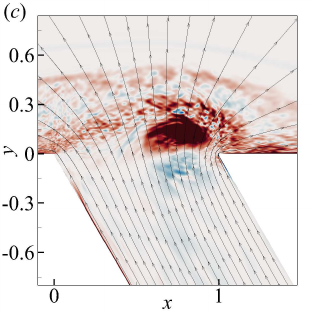}};
		\begin{scope}[x={(image.south east)},y={(image.north west)}]
			\fill [white] (0.0,0.9) rectangle (0.09,1.0);
			\node[anchor=north west, inner sep=0pt] at (0,1) {(\emph{c})};
		\end{scope}
	\end{tikzpicture}
	\begin{tikzpicture}
		\node[anchor=south west,inner sep=0] (image) at (0,0) {\includegraphics[width=0.3\textwidth]{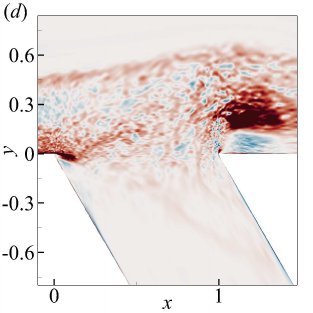}};
		\begin{scope}[x={(image.south east)},y={(image.north west)}]
			\fill [white] (0.0,0.9) rectangle (0.09,1.0);
    		\fill [white] (0.0,0.45) rectangle (0.075,0.55);
			\node[anchor=north west, inner sep=0pt] at (0,1) {(\emph{d})};
		\end{scope}
	\end{tikzpicture}
	\caption{The time evolution of the Lamb vector ($\boldsymbol{\omega} \times \boldsymbol{u}$) in the vertical direction for the $\alpha=60^{\circ}$ inclined cavity is examined for a single acoustic cycle. The contour plots capture two key time instants: those (\emph{a}) when the instantaneous acoustic source power, $\Pi$, reaches its minimum, and (\emph{c}) when it reaches its maximum. The plots (\emph{b}) and (\emph{d}) indicate the time junctures when the instantaneous acoustic source power becomes zero, such as when $\Pi=0$. Here, the superimposed streamline represents the instantaneous acoustic particle velocity field.}
	\label{fig:Fig21}
\end{figure}
Figure \ref{fig:Fig20} shows the time evolution of the Lamb vector and the acoustic particle velocity field in an $\alpha=30^{\circ}$ inclined cavity over an acoustic cycle. The acoustic absorption phase is depicted in figures \ref{fig:Fig20}(a,b), where hydrodynamic instabilities within the shear-layer absorb acoustic energy to form a coherent vortex near the upstream corner, while the residual vorticity near the downstream corner from the preceding cycle also contributes to this absorption. The second half of the acoustic production phase is captured in figures \ref{fig:Fig20}(c,d), where the vorticity-dominated regions are now in phase with the acoustic particle velocity field until their ejection from the cavity in the same direction as the acoustic particle velocity. Subsequently, figure \ref{fig:Fig21} shows the time evolution of the Lamb vector and the acoustic particle velocity field in the $\alpha=60^{\circ}$ inclined cavity over an similar acoustic cycle. Although there are significant similarities with the $\alpha=30^{\circ}$ inclined cavity, two key differences are observed. First, the vortex structure in the $\alpha=60^{\circ}$ inclined cavity exhibits enhanced spanwise coherence, contributing to a more pronounced Lamb vector and consequently enhanced instantaneous acoustic source power, according to Eq. (\ref{eq:HEC}). Second, the diminished residual vorticity near the downstream corner in the $\alpha=60^{\circ}$ inclined cavity reduces the overall absorption of acoustic energy. These factors may potentially explain the stronger acoustic response observed in the $\alpha=60^{\circ}$ inclined cavity.

\begin{figure}[!h]
	\centering
	\begin{tikzpicture}
		\node[anchor=south west,inner sep=0] (image) at (0,0) {\includegraphics[width=0.30\textwidth]{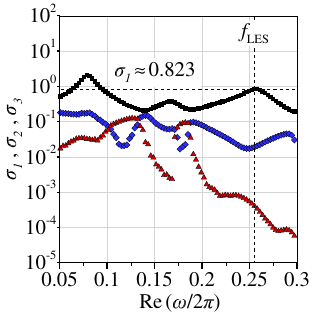}};
		\begin{scope}[x={(image.south east)},y={(image.north west)}]
			\fill [white] (0.0,0.9) rectangle (0.09,1.0);
			\node[anchor=north west, inner sep=0pt] at (0,1) {(\emph{a})};
		\end{scope}
	\end{tikzpicture}
	\begin{tikzpicture}
		\node[anchor=south west,inner sep=0] (image) at (0,0) {\includegraphics[width=0.30\textwidth]{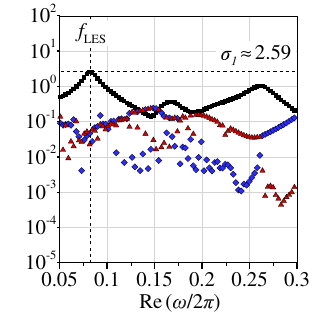}};
		\begin{scope}[x={(image.south east)},y={(image.north west)}]
			\fill [white] (0.0,0.9) rectangle (0.09,1.0);
			\node[anchor=north west, inner sep=0pt] at (0,1) {(\emph{b})};
		\end{scope}
	\end{tikzpicture}
	\begin{tikzpicture}
		\node[anchor=south west,inner sep=0] (image) at (0,0) {\includegraphics[width=0.30\textwidth]{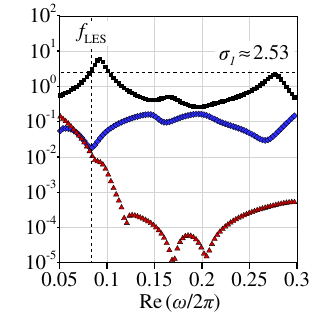}};
		\begin{scope}[x={(image.south east)},y={(image.north west)}]
			\fill [white] (0.0,0.9) rectangle (0.09,1.0);
			\node[anchor=north west, inner sep=0pt] at (0,1) {(\emph{c})};
		\end{scope}
	\end{tikzpicture}
	\caption{The low-rank behavior of the APE-resolvent operator is visualized through the first three leading magnification rates: (\protect\filledbox) $\sigma_1$, (\protect\filleddiamondblue) $\sigma_2$ and (\protect\filledtrianglered) $\sigma_3$ for (\emph{a}) $\alpha=90^{\circ}$, (\emph{b}) $60^{\circ}$ and (\emph{c}) $30^{\circ}$, respectively. The vertical dashed line (\dashed) indicates the tonal frequency observed in the LES, while the horizontal dashed line (\dashed) represents the leading amplification rate $\sigma_1$ corresponding to that frequency.}
	\label{fig:Fig22}
\end{figure}
\begin{figure}[!h]
	\centering
\begin{tikzpicture}
	\node[anchor=south west,inner sep=0] (image) at (0,0) {\includegraphics[width=0.285\textwidth]{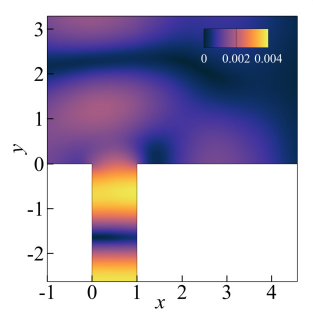}};
	\begin{scope}[x={(image.south east)},y={(image.north west)}]
		\fill [white] (0.0,0.9) rectangle (0.09,1.0);
		\node[anchor=north west, inner sep=0pt] at (0,1) {(\emph{a})};
	\end{scope}
\end{tikzpicture}
\begin{tikzpicture}
	\node[anchor=south west,inner sep=0] (image) at (0,0) {\includegraphics[width=0.285\textwidth]{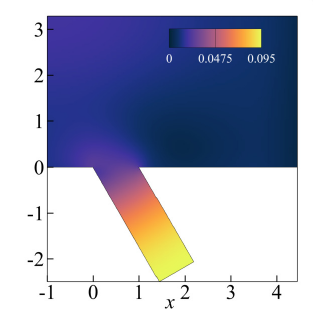}};
	\begin{scope}[x={(image.south east)},y={(image.north west)}]
		\fill [white] (0.0,0.9) rectangle (0.09,1.0);
		\node[anchor=north west, inner sep=0pt] at (0,1) {(\emph{b})};
	\end{scope}
\end{tikzpicture}
\begin{tikzpicture}
	\node[anchor=south west,inner sep=0] (image) at (0,0) {\includegraphics[width=0.285\textwidth]{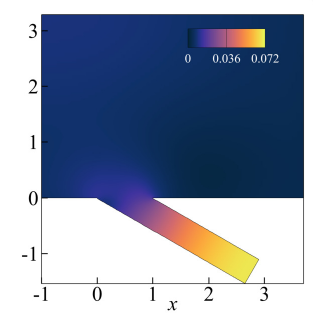}};
	\begin{scope}[x={(image.south east)},y={(image.north west)}]
		\fill [white] (0.0,0.9) rectangle (0.09,1.0);
		\node[anchor=north west, inner sep=0pt] at (0,1) {(\emph{c})};
	\end{scope}
\end{tikzpicture}
\begin{tikzpicture}
	\node[anchor=south west,inner sep=0] (image) at (0,0) {\includegraphics[width=0.285\textwidth]{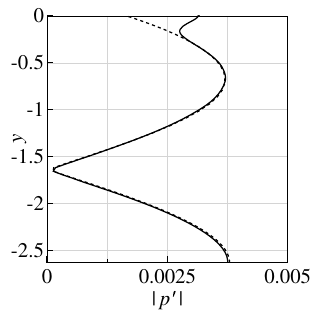}};
	\begin{scope}[x={(image.south east)},y={(image.north west)}]
		\fill [white] (0.0,0.9) rectangle (0.09,1.0);
	\end{scope}
\end{tikzpicture}
\begin{tikzpicture}
	\node[anchor=south west,inner sep=0] (image) at (0,0) {\includegraphics[width=0.285\textwidth]{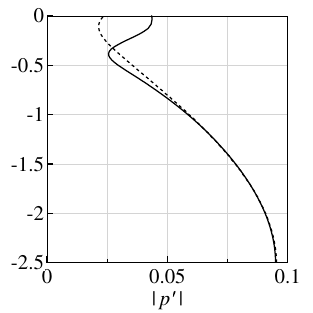}};
	\begin{scope}[x={(image.south east)},y={(image.north west)}]
		\fill [white] (0.0,0.9) rectangle (0.09,1.0);
	\end{scope}
\end{tikzpicture}
\begin{tikzpicture}
	\node[anchor=south west,inner sep=0] (image) at (0,0) {\includegraphics[width=0.285\textwidth]{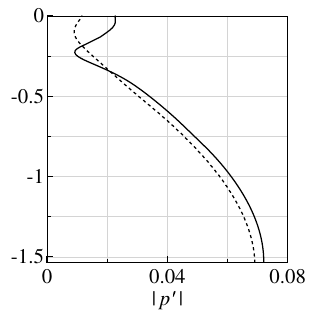}};
	\begin{scope}[x={(image.south east)},y={(image.north west)}]
		\fill [white] (0.0,0.9) rectangle (0.09,1.0);
	\end{scope}
\end{tikzpicture}
    \caption{The first row of contour plots shows the spatial distribution of the magnitude of the rank-1 reconstructed acoustic pressure field $|p'_{\text{APE}}|$ computed using Eq. (\ref{eq:1rank}) for (\emph{a}) $\alpha=90^{\circ}$, (\emph{b}) $60^{\circ}$ and (\emph{c}) $30^{\circ}$, respectively. The second row of line plots compares the depthwise distribution of wall-pressure fluctuations measured along the upstream cavity wall ($x=0$) from the (\full) LES and (\dashed) rank-1 reconstructed acoustic pressure field.}
	\label{fig:Fig23}
\end{figure}

The discussion above qualitatively explains the intense acoustic response observed in inclined cavities. To further investigate the fundamental mechanisms driving noise amplification in these cavity flow systems, an aeroacoustic resolvent analysis, as outlined in Subsection \ref{sec:ape}, has been conducted. The primary objective of this analysis is to establish a direct quantitative relationship between the Lamb vector and the magnitude of the acoustic response by examining the corresponding amplification rates and the forcing-response mode shapes of the APE-resolvent operator. Accordingly, figure \ref{fig:Fig22} presents the three leading amplification rates of the APE-resolvent operator across various frequencies. The results reveal that the leading amplification rate is significantly higher than the second and third rates at tonal frequencies across all cavity inclinations, underscoring the low-rank nature of the cavity oscillations examined in this study. Subsequently, this low-rank behavior justifies a rank-1 approximation of the APE-resolvent operator, where the dominant leading mode alone sufficiently characterizes the cavity system's acoustic response. Specifically,
\begin{equation}
	\boldsymbol{R}(\bar{q};\omega) = \hat{\boldsymbol{U}}\boldsymbol{\Sigma}\hat{\boldsymbol{V}}^{H} = \sum_i \sigma_{i} \boldsymbol{u}_{i} \boldsymbol{v}^{H}_{i} \approx \sigma_{1} \boldsymbol{u}_{1} \boldsymbol{v}^{H}_{1},
    \label{eq:aperesolventrank1}
\end{equation}
where $\boldsymbol{u}_{1}$ and $\boldsymbol{v}_{1}$ represent the leading forcing and response modes, respectively, while $\sigma_{1}$ denotes the gain of the leading forcing-response pair. The superscript $H$ in Eq. (\ref{eq:aperesolventrank1}) refers to the Hermitian transpose operation. Note that the response and forcing mode shapes are obtained from singular value decomposition, through which the leading mode shapes are normalized by construction such that $||\boldsymbol{u}_1||=||\boldsymbol{v}_1||=1$. Consequently, this approximation establishes a quantitative connection between the input forcing, $\hat{\boldsymbol{f}}_{\omega}$ (e.g., the Lamb vector), and the corresponding output acoustic field quantities, $\hat{\boldsymbol{q}}_{\omega}$ (e.g., the acoustic pressure fluctuations), which can be represented as
\begin{equation}\label{eq:1rank}
	\hat{\boldsymbol{q}}_{\omega} = \boldsymbol{R}(\bar{q};\omega) \hat{\boldsymbol{f}}_{\omega} \approx \sigma_{1} \boldsymbol{u}_{1} \boldsymbol{v}^{H}_{1}\hat{\boldsymbol{f}}_{\omega} = \sigma_{1} \boldsymbol{u}_{1} \sum_j v^*_{1,j}\hat{f}_{\omega,j} = \sigma_{1} \boldsymbol{u}_{1} F_{\omega},
\end{equation}
where the index ``\emph{j}'' denotes the grid points. Here, the measure of the source and sink is the correlation between the leading hydrodynamic forcing ($\hat{f}_{\omega,j}$) and the acoustic response ($v^{*}_{1,j}$), i.e. $v^*_{1,j}\hat{f}_{\omega,j}$ at a given point in space (\emph{j}). If the correlation is positive (in either of the $x$- or $y-$direction), the location can be regarded as a source region ($v^*_{1,j}\hat{f}_{\omega,j}>0$), and if negative, a sink region ($v^*_{1,j}\hat{f}_{\omega,j}<0$). Meanwhile, $F_{\omega} = \boldsymbol{v}^{H}_{1}\hat{\boldsymbol{f}}_{\omega}=\boldsymbol{v}^{*}_{1}\cdot\hat{\boldsymbol{f}}_{\omega}$ is the sum of all sources and sinks. Figure \ref{fig:Fig23} presents the spatial distribution of the reconstructed acoustic pressure field calculated using Eq. (\ref{eq:1rank}), which shows strong agreement with the LES data, thereby reconfirming the suitability of the rank-1 approximation for capturing the dominant acoustic response of cavity flows examined in this study.

\begin{figure}
	\centering
\begin{tikzpicture}
	\node[anchor=south west,inner sep=0] (image) at (0,0) {\includegraphics[width=0.30\textwidth]{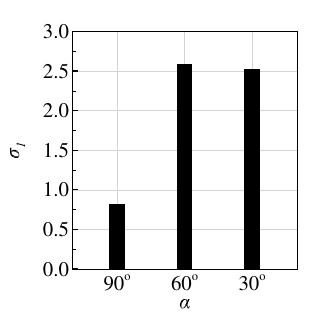}};
	\begin{scope}[x={(image.south east)},y={(image.north west)}]
		\fill [white] (0.0,0.9) rectangle (0.09,1.0);
		\node[anchor=north west, inner sep=1pt] at (0,1) {(\emph{a})};
	\end{scope}
\end{tikzpicture}
\begin{tikzpicture}
	\node[anchor=south west,inner sep=0] (image) at (0,0) {\includegraphics[width=0.30\textwidth]{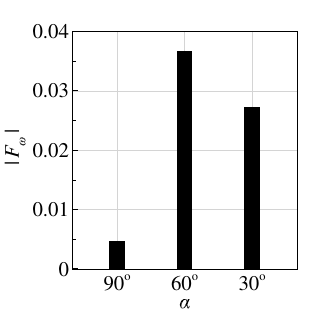}};
	\begin{scope}[x={(image.south east)},y={(image.north west)}]
		\fill [white] (0.0,0.9) rectangle (0.09,1.0);
		\node[anchor=north west, inner sep=1pt] at (0,1) {(\emph{b})};
	\end{scope}
\end{tikzpicture}
\begin{tikzpicture}
	\node[anchor=south west,inner sep=0] (image) at (0,0) {\includegraphics[width=0.30\textwidth]{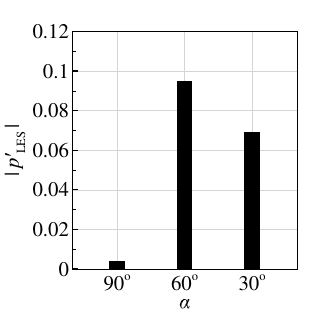}};
	\begin{scope}[x={(image.south east)},y={(image.north west)}]
		\fill [white] (0.0,0.9) rectangle (0.09,1.0);
		\node[anchor=north west, inner sep=1pt] at (0,1) {(\emph{c})};
	\end{scope}
\end{tikzpicture}
    \caption{Histogram plots showing (\emph{a}) the first leading gain of the APE-resolvent operator $\sigma_{1}$; (\emph{b}) the magnitude of the sum of sources and sinks $|F_{\omega}|$; and, (\emph{c}) the magnitude of the acoustic pressure $|p'_{\text{LES}}|$, for three different cavity inclinations.}
 \label{fig:Fig24}
\end{figure}

\begin{figure}
	\centering
\begin{tikzpicture}
	\node[anchor=south west,inner sep=0] (image) at (0,0) {\includegraphics[width=0.285\textwidth]{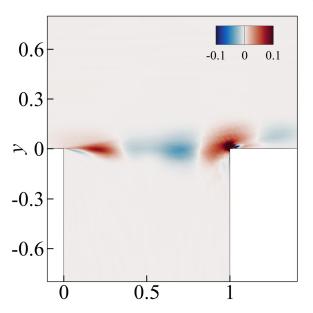}};
	\begin{scope}[x={(image.south east)},y={(image.north west)}]
		\fill [white] (0.0,0.9) rectangle (0.09,1.0);
		\node[anchor=north west, inner sep=1pt] at (0,1) {(\emph{a})};
	\end{scope}
\end{tikzpicture}
\begin{tikzpicture}
	\node[anchor=south west,inner sep=0] (image) at (0,0) {\includegraphics[width=0.285\textwidth]{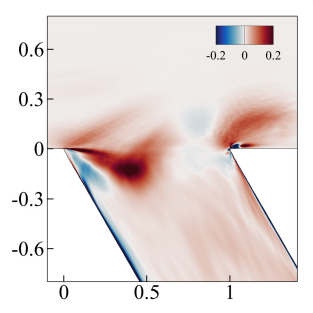}};
	\begin{scope}[x={(image.south east)},y={(image.north west)}]
		\fill [white] (0.0,0.9) rectangle (0.09,1.0);
		\node[anchor=north west, inner sep=1pt] at (0,1) {(\emph{b})};
	\end{scope}
\end{tikzpicture}
\begin{tikzpicture}
	\node[anchor=south west,inner sep=0] (image) at (0,0) {\includegraphics[width=0.285\textwidth]{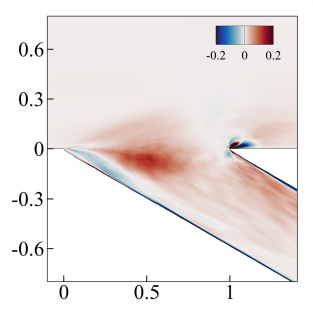}};
	\begin{scope}[x={(image.south east)},y={(image.north west)}]
		\fill [white] (0.0,0.9) rectangle (0.09,1.0);
		\node[anchor=north west, inner sep=1pt] at (0,1) {(\emph{c})};
	\end{scope}
\end{tikzpicture}
    \hspace{5cm}
\begin{tikzpicture}
	\node[anchor=south west,inner sep=0] (image) at (0,0) {\includegraphics[width=0.285\textwidth]{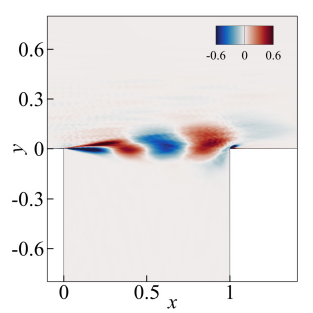}};
	\begin{scope}[x={(image.south east)},y={(image.north west)}]
		\fill [white] (0.0,0.9) rectangle (0.09,1.0);
	\end{scope}
\end{tikzpicture}
\begin{tikzpicture}
	\node[anchor=south west,inner sep=0] (image) at (0,0) {\includegraphics[width=0.285\textwidth]{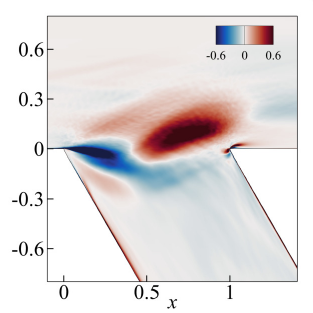}};
	\begin{scope}[x={(image.south east)},y={(image.north west)}]
		\fill [white] (0.0,0.9) rectangle (0.09,1.0);
	\end{scope}
\end{tikzpicture}
\begin{tikzpicture}
	\node[anchor=south west,inner sep=0] (image) at (0,0) {\includegraphics[width=0.285\textwidth]{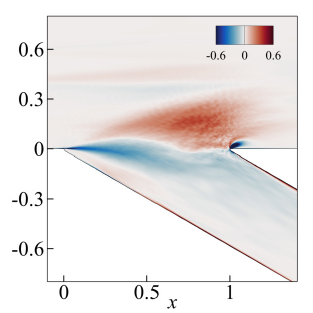}};
	\begin{scope}[x={(image.south east)},y={(image.north west)}]
		\fill [white] (0.0,0.9) rectangle (0.09,1.0);
	\end{scope}
\end{tikzpicture}
    \caption{The spatial distribution of the real part of $\boldsymbol{v}^{*}_{1} \odot \hat{\boldsymbol{f}}_{\omega}$ in the streamwise direction (top panels) and vertical direction (bottom panels) for (\emph{a}) $\alpha=90^{\circ}$, (\emph{b}) $60^{\circ}$ and (\emph{c}) $30^{\circ}$. Here, the sources and sinks are represented by regions where $\boldsymbol{v}^{*}_{1} \odot \hat{\boldsymbol{f}}_{\omega} > \boldsymbol{0}$ and $\boldsymbol{v}^{*}_{1} \odot \hat{\boldsymbol{f}}_{\omega} < \boldsymbol{0}$, respectively. The phase of the plots is selected such that the imaginary part of $F_{\omega}$ is zero.}
	\label{fig:Fig25}
\end{figure}

Figure \ref{fig:Fig24} provides useful information, showing that $\sigma_1$ and $F_\omega$ obtained from the APE-resolvent operator correlate well with the sound pressure levels directly observed from the current LES ($p'_\text{LES}$) for different cavity inclinations. First, the leading gains $\sigma_1$ are higher in the inclined cavities because they are driven by the first depthwise acoustic modes, which exhibit smaller radiation losses (i.e., more perturbation energy contained within the cavity) than those in the orthogonal cavity, as previously discussed in figure \ref{fig:Fig9}(\emph{b}). Second, the magnitude of the sum of all sources and sinks $|F_{\omega}|$ also appears greater in the inclined cavities. By contrast, the orthogonal cavity seems to suffer a significant source-sink cancellation, perhaps related to the higher hydrodynamic and acoustic modes that prevailed. Figure \ref{fig:Fig25}(\emph{a}) shows source ($v^*_{1,j}\hat{f}_{\omega,j}>0$) and sink regions ($v^*_{1,j}\hat{f}_{\omega,j}<0$) in the orthogonal cavity case where the two opposite regions have an almost equal size and magnitude leading to a mutual cancellation. Conversely, the inclined cavities exhibit unequal sizes of the source and sink regions indicating less effective cancellation between the two, as shown in figures \ref{fig:Fig25}(\emph{b}) and (\emph{c}). This reduced source-sink cancellation, together with the lower radiation losses of the first depthwise mode, directly leads to the higher sound‐pressure levels observed in inclined cavities.

\begin{figure}[!h]
	\centering
	\includegraphics[width=0.45\textwidth]{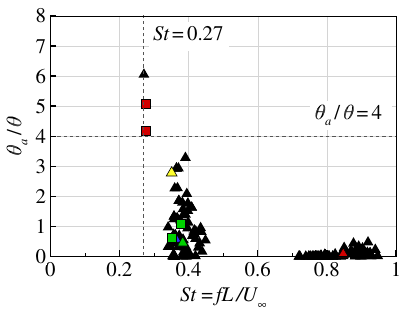}
	\caption{The scatter plot shows the ratio of acoustic particle displacement $\theta_a$ to the momentum thickness $\theta$ at three distinct flow-acoustic resonance regimes. Different symbol colors represent data from various studies on orthogonal cavity flows: (\protect\filledtriangleblack) Yang et al. \cite{Yang2009}, (\protect\filledtriangleyellow) Bagwell \cite{Bagwell2006}, (\protect\filledtriangleblue) Forestier et al. \cite{Forestier2003}, (\protect\filledtrianglegreen) Ho and Kim \cite{ho2021wall}, and (\protect\filledtrianglered) represents the current LES result at $M_\infty=0.3$. The current LES results for inclined cavities at (\protect\filledboxgreen) $M_\infty=0.2$ and (\protect\filledboxred) $0.3$ are provided. For fair comparison with experimental studies, the momentum thickness $\theta$ is measured at 10\% away from the upstream cavity corner, and the values for each cavity case are provided in Table \ref{table:Table1}. The horizontal line suggests that the ratio between acoustic particle displacement and momentum thickness ($\theta_{a}/\theta \geq 4$) may serve as an indicator for predicting the onset of amplified resonance at $St= 0.27$.}
	\label{fig:Fig26}
\end{figure} 

Thus far, the discussions have primarily centered on elucidating the aeroacoustic characteristics of cavity oscillations, with limited emphasis on the influence of incoming flow properties. To address this gap, we draw on insights from the acoustic particle velocity analysis at varying pulsation levels, as discussed in Section \ref{sec:pfluc}, along with boundary layer information presented in Table \ref{table:Table1}, to assess the impact of acoustic forcing and boundary layer thickness near the upstream corner on the cavity oscillation frequency. Here, the effect of acoustic forcing is quantified by evaluating the peak acoustic particle displacement induced by depthwise acoustic resonance, expressed as
\begin{equation}\label{eq:moma_mom}
	\theta_{a}=\frac{|p'|}{\rho_{\infty} a_{\infty} 2\pi He},
\end{equation}
where $|p'|$ denotes the magnitude of the Fourier-transformed pressure fluctuations measured at the cavity base, and $He$ represents the peak Helmholtz number identified in the pressure spectra of the LES results. According to Bagwell \cite{Bagwell2006}, lock-in oscillations may occur when the acoustic particle displacement is on the same order of magnitude as the momentum thickness. Subsequently, we compare our LES data with previous experimental findings on deep cavity flows to provide further insight into the relationship between acoustic particle displacement, momentum thickness and cavity oscillation dynamics.

Figure \ref{fig:Fig26} presents the ratios of acoustic particle displacement to momentum thickness ($\theta_{a}/\theta$) as a function of the peak Strouhal number collected from previous experimental data and the authors' computational work. The plot reveals three distinct flow-acoustic resonance regimes, i.e. a low-, mid- and high-frequency regimes centred around $St\sim 0.27$, $0.4$ and $0.85$, respectively. It is worth noting that, apart from the present cases, all data points are derived from orthogonal deep cavity configurations. Accordingly, a particularly compelling observation is the presence of a single experimental case reported by Yang et al. \cite{Yang2009} which also occurred at $St=0.27$ (coinciding with that observed in the current study of inclined cavities) despite it originated from an orthogonal cavity. Yang et al. reported it as an unexpected and intense cavity resonance mode referred to as ``$h$1*$a$1''. Although data around $St \sim 0.27$ are limited, we can see a convincing trend that a higher value of $\theta_{a}/\theta$ is required to produce a resonance at a lower frequency mode. While it is premature to draw a conclusion with the small number of samples available to date, we suggest that a threshold condition $\theta_a/\theta>4$ may be used to predict a low-frequency deep-cavity resonance at $St\sim 0.27$.

\section{Concluding remarks}\label{sec:conclusion}
We investigated the aeroacoustic behavior of deep cavities with an aspect ratio of $D/L = 2.632$ and three different inclination angles ($\alpha=30^\circ$, $60^\circ$, and $90^\circ$) at two different Mach numbers ($M_\infty=0.2$ and $0.3$) using wall-resolved large-eddy simulations. The inclined cavities at $M_\infty=0.3$ generated unexpected acoustic responses with peak amplitudes nearly 30 dB higher than those observed with the orthogonal cavity. Moreover, the peak frequency ($St=0.276$) was significantly lower compared to the orthogonal case ($St=0.849$). This was not predicted by Rossiter's model as that accounts solely for the streamwise feedback mechanism. Various analysis methods were used to investigate responsible physical mechanisms that generated the unexpected results from the inclined cavities. For the orthogonal cavity, a lock-in event occurred between the 2nd depthwise acoustic mode ($He\approx 0.255$) and the 2nd hydrodynamic mode ($St=0.849$) which exhibited two small vortices travelling across the cavity opening simultaneously. Coincidentally, this frequency also matched Rossiter's prediction with $St_n=(n-1/4)/(M_\infty+1/\kappa)$ where $n=2$ and $\kappa=0.57$. In contrast, the inclined cavities resulted in a lock-in between the 1st depthwise acoustic mode ($He=0.083$) and the 1st hydrodynamic mode ($St=0.276$) which the Rossiter's model did not predict. The 1st hydrodynamic mode which involves only one vortex across the cavity opening was due to a significantly different vortex dynamics produced in the inclined cavities. The identified vortex dynamics consists of a pronounced level of Kelvin-Helmholtz instability in the shear layer that produces a roll-up vortex that spends a substantial amount of time growing by merging smaller eddies rather than consistently travelling downstream. This merging process slowed the vortex convection speed and prolonged the vortex dwell time, yielding a lower resonance frequency locked to the 1st acoustic mode.

We suggest that the enhanced Kelvin-Helmholtz instability observed in the inclined cavities may be linked to the reduced radiation losses of the acoustic mode identified through the acoustic modal analysis. Specifically, the lower radiation loss (relative to the orthogonal-cavity case) allows more acoustic perturbation energy to remain within the cavity, which in turn more effectively amplifies the shear-layer oscillations. We quantified this effect by tracking changes in mass flow rate across the cavity opening and by examining streamline displacements. Subsequently, aeroacoustic resolvent analysis revealed the source and sink regions in each cavity, showing that source-sink cancellation was less pronounced in the inclined cavity configurations. This reduced cancellation, combined with the lower radiation losses of the 1st depthwise mode, directly led to the higher sound-pressure levels observed in the inclined cavities. Finally, we hypothesize that the ratio between the acoustic particle displacement and the momentum thickness $\theta_{a}/\theta>4$ might be a necessary condition for the onset of the low-frequency resonance ($St\sim 0.27$) in deep and inclined cavities.

While the current study provides valuable insights into flow-acoustic resonances in deep and inclined cavity configurations, it has certain limitations that warrant further research. First, the findings are specific to the aspect ratio of $L/D = 2.632$ and the Mach numbers of 0.2 and 0.3 considered in this study. Further investigations are required to better understand the distinctive vortex dynamics and the low-frequency mode selection process across various aspect ratios and over a broader range of Mach and Reynolds numbers. Second, the effects of three-dimensional geometries such as cavities with a finite span or circular cross-sections with various inclinations remains poorly understood. Third, variations in cavity-floor configurations (for example, non-orthogonal side walls or non-planar floors) may alter resonance frequencies and amplitudes. Future work should explore optimal cavity geometries and/or flow conditions that may lead to strategies for mitigating or controlling deep cavity resonance, given the significant practical implications for various engineering applications.

\begin{acknowledgments}
The authors would like to acknowledge the studentship provided by Rolls-Royce UTC (University Technology Centre) for Propulsion Systems Noise at the University of Southampton. The authors also thank EPSRC (Engineering and Physical Sciences Research Council) for the computational time made available on the UK supercomputing facility ARCHER2 via the UK Turbulence Consortium (EP/R029326/1).
\end{acknowledgments}

\bibliography{apssamp}

\end{document}